\documentclass[apj]{emulateapj}
\usepackage{apjfonts}

\shorttitle{SUPERMODEL ANALYSIS OF GALAXY CLUSTERS}
\shortauthors{FUSCO-FEMIANO, CAVALIERE \& LAPI}
\journalinfo{Accepted by ApJ.}

\begin{document}

\title{Supermodel Analysis of Galaxy Clusters}
\author{R. Fusco-Femiano\altaffilmark{1}, A. Cavaliere\altaffilmark{2}, A.
Lapi\altaffilmark{2,3}}\altaffiltext{1}{INAF $-$ Istituto di
Astrofisica Spaziale e Fisica Cosmica, Via Fosso del Cavaliere,
00133 Roma, Italy.}\altaffiltext{2}{Dip. Fisica, Univ. `Tor
Vergata', Via Ricerca Scientifica 1, 00133 Roma, Italy.}
\altaffiltext{3}{Astrophysics Sector, SISSA/ISAS, Via Beirut
2-4, 34151 Trieste, Italy.}

\begin{abstract}
We present the analysis of the X-ray brightness and temperature
profiles for six clusters belonging to both the Cool Core and
Non Cool Core classes, in terms of the Supermodel (SM)
developed by Cavaliere, Lapi \& Fusco-Femiano (2009). Based on
the gravitational wells set by the dark matter halos, the SM
straightforwardly expresses the equilibrium of the IntraCluster
Plasma (ICP) modulated by the entropy deposited at the boundary
by standing shocks from gravitational accretion, and injected
at the center by outgoing blastwaves from mergers or from
outbursts of Active Galactic Nuclei. The cluster set analyzed
here highlights not only how simply the SM represents the main
dichotomy Cool vs. Non Cool Core clusters in terms of a few ICP
parameters governing the radial entropy run, but also how
accurately it fits even complex brightness and temperature
profiles. For Cool Core clusters like A2199 and A2597, the SM
with a low level of central entropy straightforwardly yields
the characteristic peaked profile of the temperature marked by
a decline toward the center, without requiring currently strong
radiative cooling and high mass deposition rates. Non Cool Core
clusters like A1656 require instead a central entropy floor of
a substantial level, and some like A2256 and even more A644
feature structured temperature profiles that also call for a
definite floor extension; in such conditions the SM accurately
fits the observations, and suggests that in these clusters the
ICP has been just remolded by a merger event, in the way of a
remnant cool core. The SM also predicts that dark matter halos
with high concentration should correlate with flatter entropy
profiles and steeper brightness in the outskirts; this is
indeed the case with A1689, for which from X rays we find
concentration values $c\sim 10$, the hallmark of an early halo
formation. Thus we show the SM to constitute a fast tool not
only to provide wide libraries of accurate fits to X-ray
temperature and density profiles, but also to retrieve from the
ICP archives specific information concerning the physical
histories of dark matter and baryons in the inner and the outer
cluster regions.
\end{abstract}

\keywords{galaxies: clusters: general --- galaxies: clusters:
individual (A2199, A2597, A1689, A1656, A2256, A644) ---
X-rays: galaxies: clusters}

\section{Introduction}

The evolution of a galaxy cluster comprises relaxed stretches
punctuated by violent merger events affecting the
gravitationally dominant dark matter (DM) halo, especially
frequent in its early life after the initial collapse.

A number of such events may be caught in action, as in the
extreme case of the Bullet Cluster (Markevitch et al. 2002;
Clowe et al. 2006). As to the many quieter clusters, one may
wonder how long and how precise a memory of similar if less
dramatic events --- and of the more frequent ones caused by
central outbursts from Active Galactic Nuclei (AGNs) --- is
retained by the hot diffuse baryons constituting their
IntraCluster Plasma (ICP). We will show how a considerable
amount of quantitative information may be retrieved even
several Gyrs after the event from the X-ray bremsstrahlung
emission by the ICP in the inner cluster regions.

On the other hand, the quiet stretches when external DM and gas
are smoothly accreted across the virial radius $R$ will shape
the outer DM density and gravitational potential, and hence the
ICP distribution and X-ray bremsstrahlung emission in the
cluster outskirts. Is there any simple and comprehensive,
physical modeling conducive to a precise yet fast analysis of
both these regions?

We will show how indeed our Supermodel (see Cavaliere, Lapi \&
Fusco-Femiano 2009, thereafter CLFF09) enables us to accurately
fit the full radial profiles of the X-ray observables, namely,
the temperature $T(r)$ and the brightness $S(r)\propto n^2(r)\,
T^{1/2}(r)$ that provides the baryon number density $n$. Whence
we read out the imprints of the \emph{thermodynamic} cluster
history in the form of level, pattern, and time for depositions
or injections of ICP specific entropy $k\equiv k_B T/n^{5/3}$
in Boltzmann units. Rather than relying on entropy profiles in
the literature, we will focus on deriving the radial entropy
runs from fitting the modeled primary observables for which
precise, resolved and robust data (typically out to $R/2$)
exist as is the case for the six clusters considered here.

The Supermodel (SM) straightforwardly expresses the profiles
$n(r)$ and $T(r)$ in the entropy-modulated equilibrium of the
ICP within the potential wells provided by the dominant DM.
These two components are related not only by their common
containing well, but also by parallel accretion of surrounding
DM and baryons into the cluster volume. Thus the SM tells the
thermal and dynamical past of a cluster back to its formation
time, to include subsequent mergers or AGN outbursts and
current outer accretion.

The SM is built upon the physical DM distributions derived by
Lapi \& Cavaliere (2009a) from the Jeans equation. These base
on the radial run of the functional $K\equiv
\sigma^2/\rho^{2/3}$, namely the `DM entropy', that combines
the 1-D velocity dispersion $\sigma(r)$ with the density
$\rho(r)$. From \emph{N}-body simulations $K(r)$ is found to
constitute a truly universal halo feature as it always follows
a powerlaw run $K(r)\propto r^{\alpha}$ throughout the halos'
bulk (Taylor \& Navarro 2001; Dehnen \& McLaughlin 2005;
Hoffman et al. 2007; Vass et al. 2009; Navarro et al. 2009). On
the other hand, Lapi \& Cavaliere (2009a) compute $\alpha$, and
find it to lie in the range $1.25 - 1.3$ from galaxies to
massive clusters, narrowed down to $1.27 - 1.3$ from poor
(overall mass $M \sim$ a few $10^{14}\,M_{\odot}$) to very rich
($M \sim$ a few $10^{15}\,M_{\odot}$) clusters. Two basic
features of these `$\alpha$-profiles' entering the SM frame are
briefly recalled in Appendix A.

We stress that our treatment of the $\alpha$-profiles (see Lapi
\& Cavaliere 2009a,b) agrees with recent numerical simulations
(Zhao et al. 2003; Diemand et al. 2007; Wang \& White 2008) in
picturing the DM halo formation as a \emph{two-stage}
development. The epoch $z_t$ ranging from $0.5$ to $2$ marks
the halo's transition from the stage of fast violent collapse
scarred by major merging events, to one of progressively
smoother and slower accretion building up the outskirts with
little body perturbation.

During the latter, the DM profiles develop an increasing
concentration $c \equiv R/r_{-2}$, that actually measures the
current outer extension out to the virial radius $R$, relative
to the region inner to $r_{-2}$ where $\rho(r)$ is flatter than
$r^{-2}$ (see CLFF09). Starting with values $c\approx 3.5$ at
the transition, $c$ increases to current values $c\approx 3.5\,
(1+z_t)$, to attain values $c\approx 10$ or more for the $10\%$
fraction of rich clusters with a transition as early as $z_t
\approx 1.5$ (see Lapi \& Cavaliere 2009b); thus $c$ measures
the \emph{dynamic} age $z_t$ of the cluster. Since $R$ is
typically of order $2$ Mpc, values $c\approx 5-10$ imply
$r_{-2}\approx 100-200$ kpc; in the following, we will refer to
`outskirts' for the region $r>r_{-2}$, and to `inner body' for
the region $r\la r_{-2}$. Based on the SM, we will discuss in
particular how the outer distribution of the ICP entropy
relates to the DM concentration.

The SM describes the equilibrium of the ICP in the DM potential
well, as we recall in \S~2; there we briefly recap the main
features of the ICP description in the form of the SM as
developed by CLFF09, with the addition of Eq.~(8) that enables
computing the total mass. Then in \S~3 we describe our analysis
procedure based on the SM. In \S~4 we apply such a procedure to
the data for a set of six clusters with diverse profiles of
X-ray brightness and temperature. Finally, in \S~5 we discuss
the specific physical information we extract from our analysis
of the X-ray observations.

In our treatment we adopt a standard flat cosmology with
normalized matter density $\Omega_M = 0.27$, dark energy
density $\Omega_{\Lambda} = 0.73$ and Hubble constant $H_0 =
70$ km s$^{-1}$ Mpc$^{-1}$, except when comparing with data
otherwise scaled.

\section{The Supermodel}

We concentrate first on DM halos close to self-gravitational
equilibrium under smooth and slow accretion, past the fast
collapse stage and after any residual violent mergers, as
discussed in \S~5. Within these DM gravitational wells, the ICP
approaches hydrostatic conditions after outgoing blastwaves and
shocks driven by central mergers or AGN outbursts have
subsided.

In such conditions, the ICP density is governed by the balance
between the gravitational force and the gradient of the
pressure $p=n\,k_BT/\mu$; with the latter expressed as $p
\propto k\, n^{5/3}$ in terms of ICP entropy, the balance reads
\begin{equation}
\frac{1}{\mu m_p n}{d\, (k \, n^{5/3}) \over dr} = -\frac{G\,
M(<r)}{r^2}~,
\end{equation}
where $\mu\approx 0.6$ is the mean molecular weight, $m_p$ is
the proton mass, and $G$ is the gravitational constant.

Once the radial entropy run $k(r)$ is given as discussed below,
the solution of this $1^{\rm st}$ order differential equation
allowed us (CLFF09) to write the profiles of the gas
temperature and density in the form
\begin{equation}
\bar{T}(\bar{r}) = \bar{k}(\bar{r})\, \bar{n}^{2/3}(\bar{r}) =
\bar{k}^{\,3/5}(\bar{r})\, \left[1+ \frac{2}{5}\, b_R\,
\int_{\bar{r}}^1 \, \mathrm{d}\bar{r}'~{\bar{v}^2_c(\bar{r}')\over
\bar{r}'}\, \bar{k}^{-3/5}(\bar{r}')\right]~.
\end{equation}
Here barred variables are normalized to their boundary value at
$r = R$; the squared circular velocity $\bar{v}^2_c(\bar{r}) =
\bar{M} (< \bar{r})/\bar{r}$ is taken from the
$\alpha$-profiles with their weak dependence on $\alpha$, and
is expressed in Appendix A in terms of hypergeometric
functions; finally, we define $b_R \equiv \mu m_p
v^2_c(R)/k_BT(R)$.

The latter incorporates the boundary condition required for
solving Eq.~(1). It is physically convenient to fix such a
reference value at the virial radius $r\approx R$ rather than at
the center, where $v_c^2(r)$ vanishes steeply (see Lapi \&
Cavaliere 2009a) while the ICP is often affected by violent
stochastic events such as mergers and AGN outbursts. Moreover,
at $r\sim R$ closely universal accretion of external
InterGalactic Medium (IGM) prevails for both Cool Core (CC) and
Non Cool Core (NCC) clusters; this holds to lowest order,
although the related conversion of infall kinetic into thermal
energy may differ somewhat related to cluster age and
preheating conditions, as discussed in \S~3 and 5.

In fact, in the absence of substantial preheating in the IGM
larger than some $1/2$ keV per particle (see Lapi et al. 2005;
McCarthy et al. 2008), the energy conversion takes place in
\emph{strong} accretion shocks that linger at the virial radius
(see Tozzi \& Norman 2001; Lapi et al. 2005; Voit 2005), to
imply
\begin{equation}
b_R = {3 \over 2\Delta \phi_{R}}~
\end{equation}
that takes on values around $2.5$. These obtain from maximal
conversion at a strong shock of the gravitational infall
energy, that yields $k_B T_R = \frac{2}{3}\mu m_p \,\Delta
\Phi_{R}$; this quantity involves the specific kinetic energy
$2\, v^2_R\Delta \phi_{R}$ gained by the IGM that starts from
the turning point its free fall toward $R$ across the potential
drop $\Delta \Phi_{R}=v_R^2\,\Delta\phi_R$ (see Lapi et al.
2005). The latter is provided by the DM $\alpha$-profiles, and
is conveniently normalized to the related $v_R^2\equiv
v_c^2(R)$; whence Eq.~(3) follows.

We stress that $k_B T_R$ lowers when $\Delta\phi$ and the
proportional infall kinetic energy are smaller owing to a large
concentration $c$; e.g., a value $c\approx 10$ holding for a
cluster with an early transition (see \S~1), in place of
$c\approx 4$ holding for clusters with a recent transition,
implies $\Delta \phi$ to decrease from $0.57$ to $0.47$ (see
CLFF09). On the other hand, $k_B T_R$ also lowers when a high
preheating level of the IGM \emph{weakens} the shock and
impairs the conversion efficiency as discussed by Lapi et al.
(2005) and Voit (2005).

To obtain the ICP temperature and density profiles the full
radial run $k(r)$ of the entropy is required. Its physical
modeling is based upon the notions that entropy is eroded by
radiative cooling on the timescale $t_c \approx 65\, (k_B T/5~
\mathrm{keV})^{1/2}\, (n/10^{-3}~\mathrm{cm}^{-3})^{-1}$ Gyr
(Sarazin 1988), whilst it is enhanced by \emph{shocks}; at
$r\approx R$ it is deposited by standing accretion shocks, and
in the central region is injected by outbound blastwaves
terminating in a shock, as are driven by supersonic outflows
from AGN outbursts or head-on mergers.

In the vicinity of $r \approx R$ the powerlaw approximation $k
\propto r^a$ always applies (see CLFF09) with boundary value
\begin{equation}
a_R = 2.4 - 0.47\, b_R~
\end{equation}
varying around $1.1$. Note that after Eq.~(3) the upper bound
$a_R = 45/19 \approx 2.4$ obtains for large $\Delta \phi$ as
then $b_R$ tends to vanish; at the other end, $a_R$ decreases
for weaker and weaker shocks corresponding to smaller
$\Delta\phi$ and/or relatively stronger preheating levels. As
no other major sources or sinks of entropy occur from the
boundary at a few Mpcs down to the central $r\sim 10^2$ kpc, in
the outer range the entropy is deposited and \emph{stratified}
during the stage of slow accretion; thus a powerlaw radial run
is set with slope $a(r)$ that stays close to its boundary value
$a_R$.

At the center, instead, entropy may be intermittently injected
by shocks driven by mergers reaching down there (McCarthy et
al. 2007; Balogh et al. 2007), and by powerful AGNs residing in
the central massive galaxies (see Valageas \& Silk 1999; Wu et
al. 2000; Scannapieco \& Oh 2004; Lapi et al. 2005) as observed
and reviewed by McNamara \& Nulsen (2007) and Markevitch \&
Vikhlinin (2007).

Thus the full entropy profile that combines central injections
with outer stratification may be described by the simple
parametric expression (see Voit 2005 and references therein)
\begin{equation}
\bar{k}(\bar{r}) = \bar{k}_c + (1 - \bar{k}_c)\, \bar{r}^a~;
\end{equation}
in fact, this approaches a constant value at small radii, and
smoothly goes into a powerlaw at large radii.

Entropy profiles similar to Eq.~(5) have been recently reported
by Cavagnolo et al. (2009) from an analysis of a
\textsl{Chandra} archival sample comprising $239$ clusters. In
fact, most of them are well fit by a power law at large radii
plus a constant value $k_0\ga k_c$ at small radii, with a
basically bimodal distribution peaked at $k_0\approx 20$ keV
cm$^{2}$ and at $k_0\approx 150$ keV cm$^{2}$.

On the other hand, some clusters show evidence of a sharper
entropy floor (e.g., Pratt et al. 2005 for A2218; see also
Fig.~5 in Cavagnolo et al. 2009) that we represent not only
with a level $k_c$ but also with a definite extension $r_{f}$,
so that the corresponding radial entropy run reads
\begin{equation}
\bar{k} = \bar{k}_c
\end{equation}
for $\bar{r}\leq \bar{r}_{f}$, and as
\begin{equation}
\bar{k} = \bar{k}_c+ (1-\bar{k}_c)\, \left({\bar{r}-\bar{r}_{f}
\over 1-\bar{r_f}}\right)^a ~
\end{equation}
for $\bar{r}>\bar{r}_{f}$. The scale $r_f$ may be interpreted
as the terminal radius just reached by an outbound blastwave
driven by a violent energy input at the center (see Lapi et al.
2005); at $r_f$ its decreasing Mach number $\mathcal{M}(r)$ has
decayed to unity and the blast has stalled and degraded into
adiabatic sound waves, as caught in action by Fabian et al.
(2006) in the Perseus Cluster (see also \S~5 for a discussion).
In this picture, Eq.~(5) represents a later stage caused by
diffusive smoothing and mixing of such a radial imprint, while
the enhanced entropy level is still high before radiative
erosion has set in.

From the gas temperature and density profiles given by Eq.~(2)
we also derive the distribution of the gravitating mass in the
form (see Sarazin 1988)
\begin{equation}
M(<\bar{r})\simeq 3.65\times 10^{13}~\left({k_B T_R\over {\rm
keV}}\right)\,\left({R\over {\rm
Mpc}}\right)\,b_R~\bar{r}~\bar{v}^2_c(\bar{r})~ M_{\odot}~,
\end{equation}
where the quantity $\bar{r}\,\bar{v}^2_c(r)$ grows slowly with
$\bar{r}$ and saturates to unity.

\begin{figure*}
\center\epsscale{0.9} \plottwo{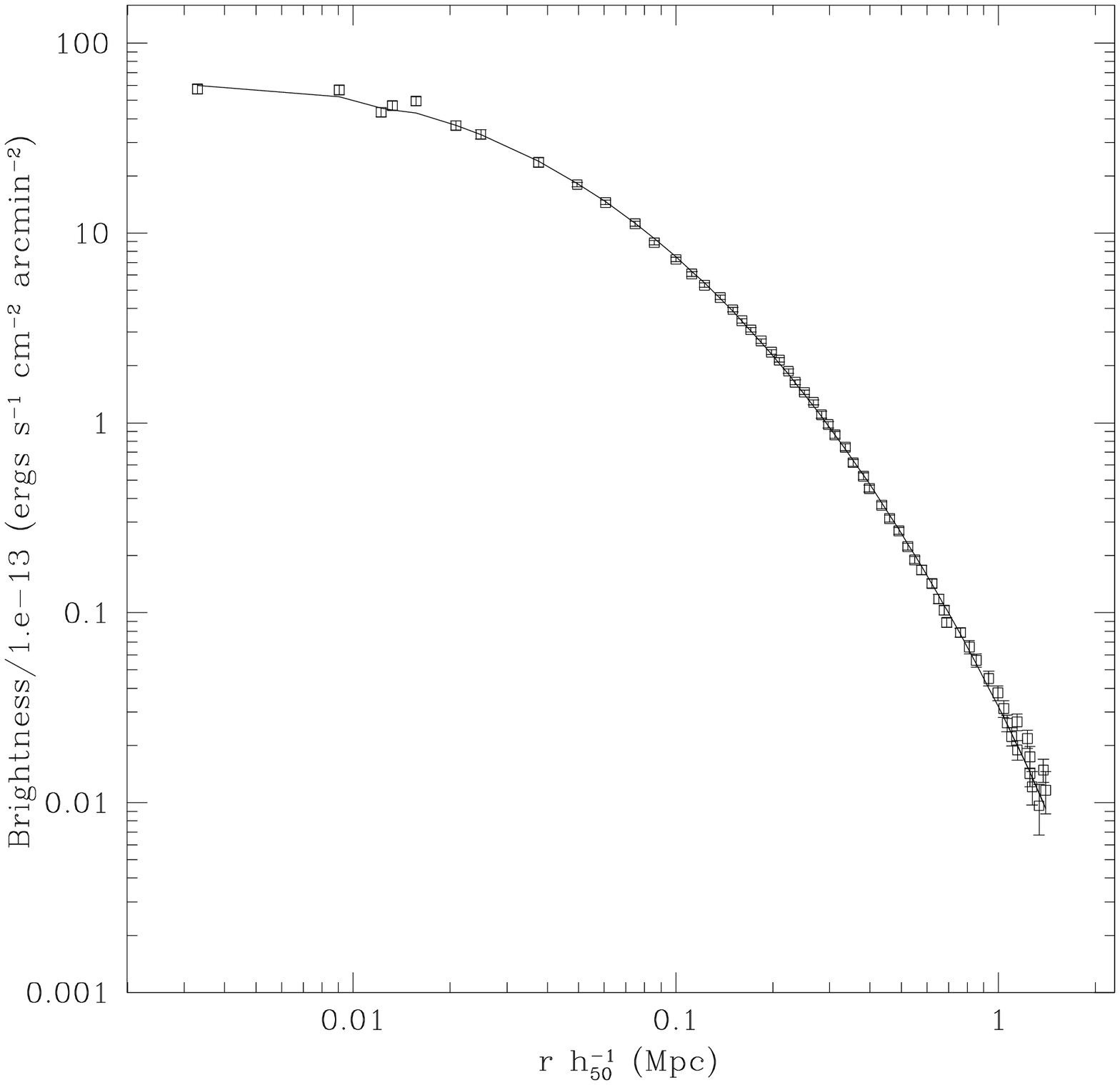}{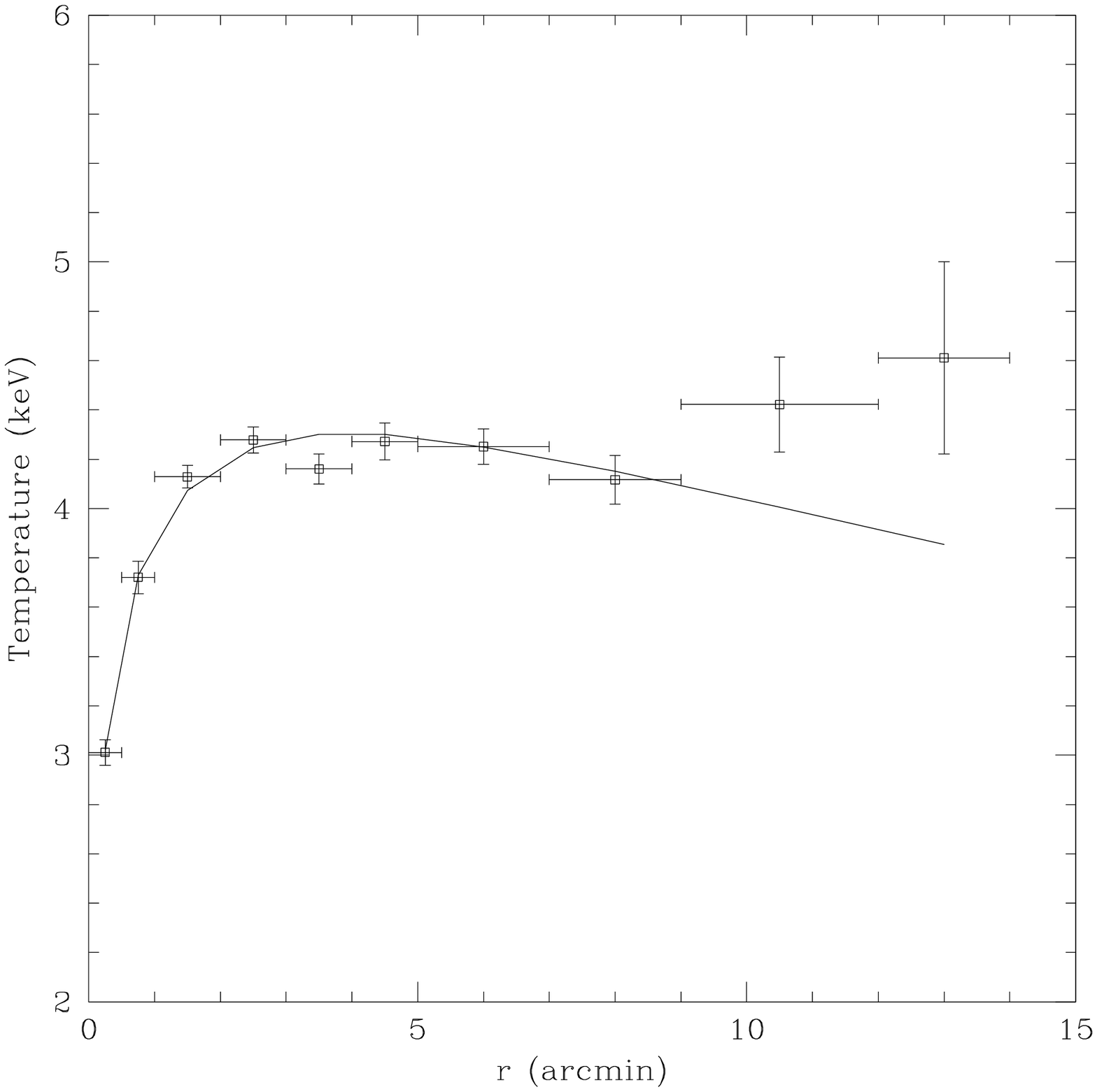} \caption{Abell
2199. {\it Left panel}: The solid line is our fit ($\chi^2=
73.9/56$) to the brightness profile measured by Mohr et al.
(1999), on adopting the entropy profile given by Eq.~(5). {\it
Right panel}: The solid line is our fit ($\chi^2$ = 15.8/5) to
the radial temperature profile measured by Snowden et al.
(2008), on adopting the entropy profile given by Eq.~(5).}
\end{figure*}

\section{Supermodel analysis}

Here we describe in detail how we use the SM to account for the
X-ray brightness and temperature profiles for both classes of
CC and NCC clusters (Molendi \& Pizzolato 2001; Leccardi \&
Molendi 2009); in particular, we will examine in detail the
following six clusters: A2199, A2597, A1689, A1656, A2256,
A644. These have been collected from the literature on the
basis of the quality and detail of the X-ray data, keeping a
balance between the CC and NCC classes. We fit with the SM the
profiles of emission weighted temperature and X-ray brightness,
as given by Eqs.~(9) and (10) below. The parameter values are
pinned down on using a standard $\chi^2$ minimization
procedure, and the uncertainties are quoted at the $68\%$
confidence level.

The free parameters are set as follows. The DM halo
distributions depend weakly on the index $\alpha$ that we fix
at the value $\alpha\approx 1.27$, and are strongly marked by
the \emph{concentration} $c$ that we leave as a free parameter.
The ICP profiles are parameterized by the \emph{slope} $a$ and
the \emph{central} value $k_c$ that define the radial entropy
run after Eq.~(5); in some clusters an acceptable fit requires
to introduce the \emph{size} $r_{f}$ of the central floor after
Eqs.~(6) and (7).

In the SM as with all models the virial radius $R$ intervenes
to set the data distance scale, as well as the bound to l.o.s.
integrations; we keep it fixed when a robust determination is
available from the literature, e.g., from observations of
galaxy dynamics, or `red sequence' termination, or
gravitational lensing. Otherwise, we determine it from fitting
(in the range where the data are reliable) the projected
profile of the emission-weighted temperature
\begin{equation}
\langle k_B T(\bar{w})\rangle =k_B T_R\,
{\int_0^{\sqrt{1-\bar{w}^2}}\,
\mathrm{d}\bar{\ell}~\bar{n}^2(\bar{r})\,\Lambda[T(\bar{r})]\,\bar{T}(\bar{r})\over
\int_0^{\sqrt{1-\bar{w}^2}}\,
\mathrm{d}\bar{\ell}~\bar{n}^2(\bar{r})\,\Lambda[T(\bar{r})]}~,
\end{equation}
in terms of the projected radius $\bar{w}\equiv w/R$. In the
following we approximate the detailed cooling function with
$\Lambda(T)\propto T^{1/2}$ as appropriate for the ICP in hot
clusters. The above relation not only pins down the
(horizontal) scale $R$, but also sets the normalization
(vertical) scale $T_R$. The knowledge of $R$ and $T_R$ allows
us to derive the ICP density $n_R$ from fitting the brightness
distribution
\begin{displaymath}
S(\bar{w})\approx S_0\, \left({R\over {\rm Mpc}}\right)\,
\left({n_R\over 10^{-3}~\mathrm{cm}^{-3}}\right)^2\,\left({k_B
T_R\over \mathrm{keV}}\right)^{1/2}\times
\end{displaymath}
\begin{equation}
\times\int_0^{\sqrt{1-\bar{w}^2}}\,
\mathrm{d}\bar{\ell}~\bar{n}^2(\bar{r})\,\bar{T}^{1/2}(\bar{r})\,
F[E_1,E_2,T(\bar r)]~;
\end{equation}
here $S_0\approx 3.4\times 10^{-13}\, (1+z)^{-4}$ erg s$^{-1}$
cm$^{-2}$ arcmin$^{-2}$, and the factor $F(E_1,E_2,T)\simeq
e^{-E_1/k_BT}-e^{-E_2/k_BT}$ takes into account specific
instrumental bands $E_2-E_1$ (e.g., Ettori 2000).

The SM actually predicts the values of $T_R$, $n_R$ and
$k_R=k_B T_R/n_R^{2/3}$ from extrapolating the profiles into
the outer cluster regions; observing the latter challenges the
sensitivity and defies the resolution of most current
instruments, but will constitute a main target for the
next-generation X-ray telescopes planned to study low surface
brightness plasmas (see \S~5).

With the use of such facilities $R$ will be read out from the
profiles as the position of the shock discontinuities in $n(r)$
and $T(r)$ above the values prevailing in the IGM (see Lapi et
al. 2005), even though such discontinuities may be blended by
complex texture of the shocks and smoothed by projection
integrations (see Tormen et al. 2004). On the other hand,
substantial preheating of the IGM may weaken the accretion
shocks and lower $T_R$. Such conditions will be indicated by
any discrepancy between the value resulting from Eq.~(9) and
the reference value $k_B T_R$ corresponding to strong shocks in
a halo of given concentration.

\section{Results for individual clusters}

Next we present the results for each of our six clusters; the
individual fitting parameters are given in the corresponding
subsections and are collected in Table~1.

\subsection{Abell 2199}

A2199 is a rich cluster at $z\approx 0.031$ centered on the cD
galaxy NGC6166 that hosts the radio source 3C338. The latter
shows evidence of several events of radio emission, as
witnessed by the clear correspondence between positions of
radio lobes and depressed X-ray surface brightness
(`cavities').

This is a CC cluster with the temperature profile declining
from about $200$ kpc toward the center as observed with
\textsl{Chandra} by Johnstone et al. (2002) and with
\textsl{XMM-Newton} by Snowden et al. (2008). Fig.~1 shows the
SM fits to the projected emission-weighted temperature profile
given by the latter authors, and to the brightness distribution
by Mohr et al. (1999) obtained with \textsl{ROSAT}. Both
profiles do not show evidence of an entropy floor down to
$r\approx 2$ kpc.

From the SM, we obtain the virial radius $R =
2.1^{+0.1}_{-0.4}$ Mpc and the halo concentration parameter
$c=6.7^{+1.0}_{-1.0}$. In the inner ICP regions we find
$\bar{k}_c = 0.39^{+0.16}_{-0.16}\times 10^{-2}$. Throughout
the cluster body we derive the entropy slope
$a=0.95_{-0.01}^{+0.01}$. In the outskirts we find $k_B T_R =
1.93^{+0.05}_{-0.05}$ keV and $n_R = 1.40^{+0.03}_{-0.03}\times
10^{-5}$ cm$^{-3}$, yielding $k_R = 3320_{-130}^{+130}$ keV
cm$^{2}$. Correspondingly, we obtain $k_c = 13^{+6}_{-6}$ keV
cm$^2$.

\begin{figure*}
\center\epsscale{0.9} \plottwo{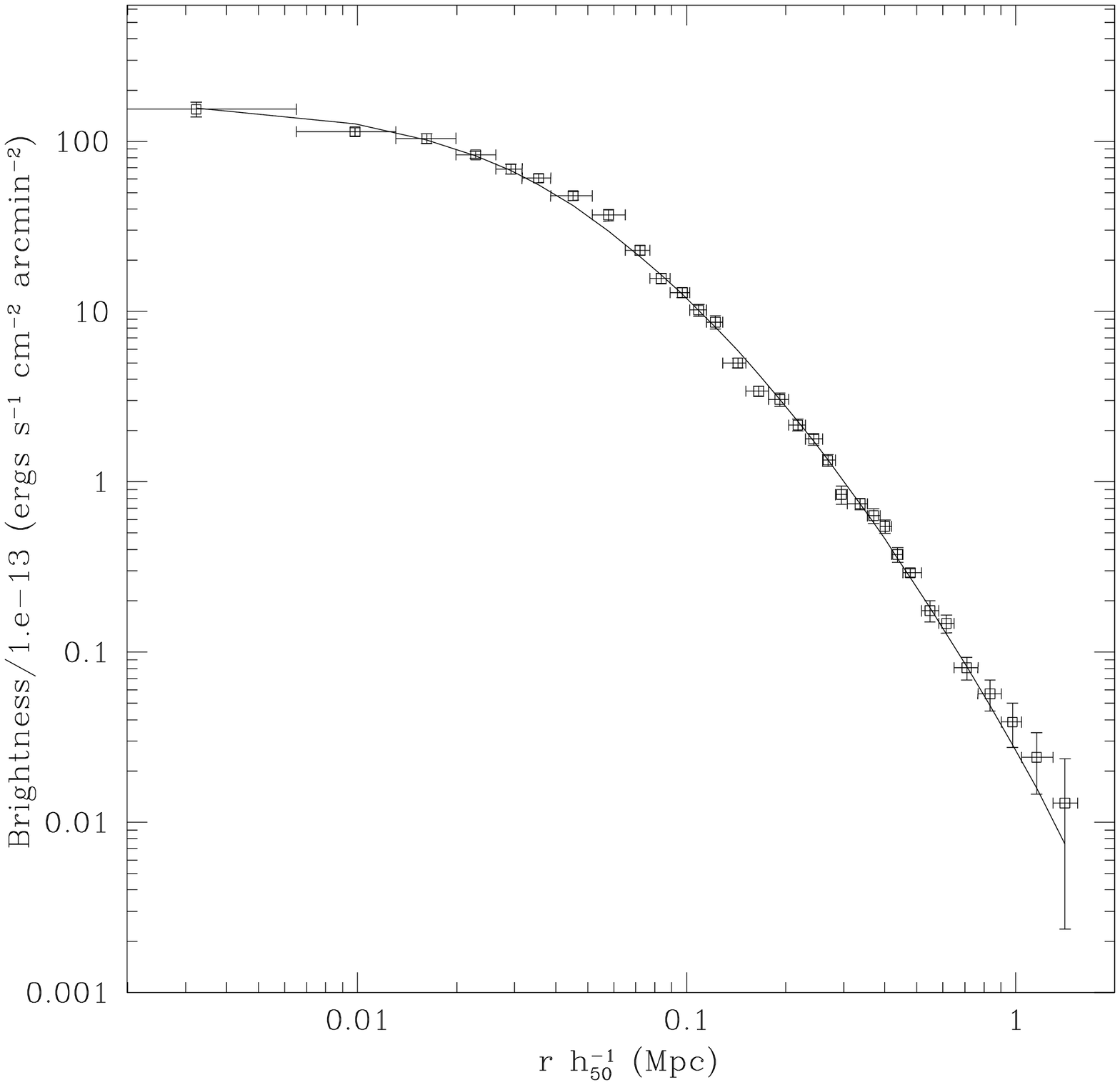}{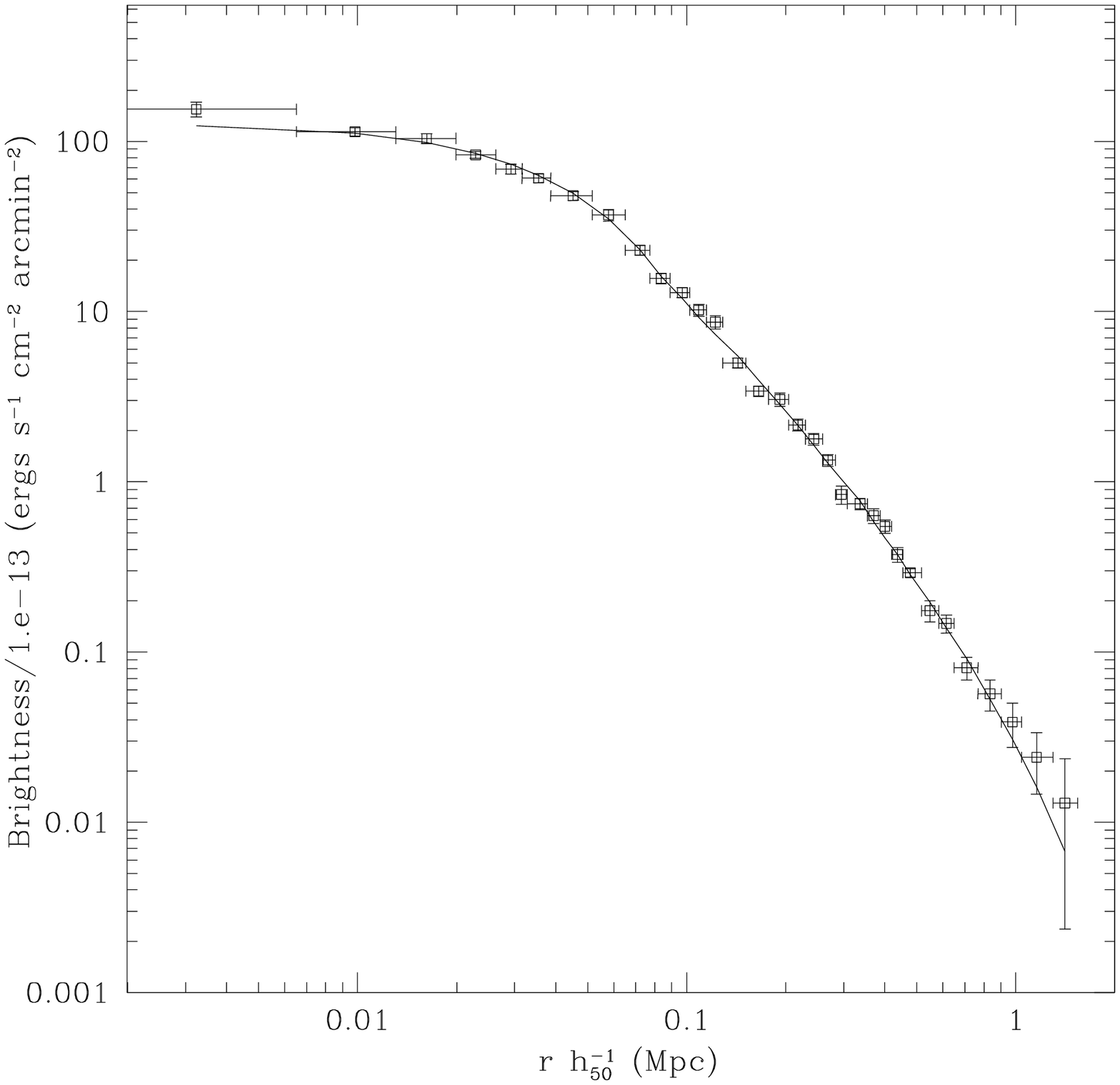}
\caption{A2597. {\it Left panel}: The solid line is our fit
($\chi^2$ = 53.5/28) to the brightness profile measured by Xue
\& Wu (2000), on adopting the entropy profile given by Eq.~(5).
{\it Right panel}: The solid line is our fit ($\chi^2$ =
32.3/27) on adopting the entropy profile given by Eqs.~(6) and
(7).}
\end{figure*}

\begin{figure}
\center\epsscale{1.}\plotone{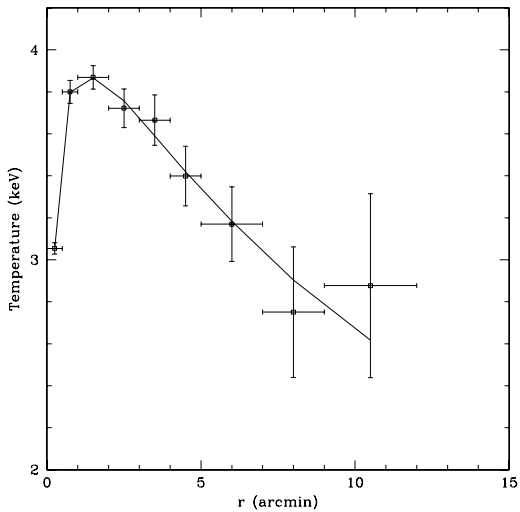}\caption{A2597. The solid
line is our fit ($\chi^2$ = 1.1/4) to the temperature profile
measured by Snowden et al. (2008), on adopting the entropy
profile given by Eq.~(5).}
\end{figure}

\subsection{Abell 2597}

A2597 is a nearby cluster at $z\approx 0.08$ that harbors the
central radio source PKS 2322-123; it shows features in
different spectral regions. In particular, \textsl{Chandra}
X-ray observations show a bright central region and two low
surface brightness ghost cavities (McNamara et al. 2001),
adding to other core structures suggestive of inner cavities
(Clarke et al. 2005). A2597 is a CC cluster as is also
confirmed with \textsl{XMM-Newton} observations by Morris \&
Fabian (2005) and Snowden et al. (2008). The average
temperature is around $2.6$ keV (Sarazin \& McNamara 1997).

We fit (see Fig.~2) the surface brightness data obtained by
\textsl{ROSAT/PSPC} observations (Xue \& Wu 2000) and the
temperature profile observed by \textsl{XMM-Newton} (Snowden et
al. 2008), on considering the two entropy profiles given by
Eqs.~(5) or by Eqs.~(6) and (7). The latter include the
additional parameter $r_f = 50^{+7}_{-7}$ kpc that turns out to
be significant at the $99.97\%$ level after the $F$-test, and
may be interpreted as the effect of the central radio source
PKS2322-123. On the other hand, the size of the entropy floor
lies within the first bin of the temperature data (see Fig.~3),
below the resolution of our profile.

From the SM, we obtain the virial radius $R =
1.9^{+0.4}_{-0.4}$ Mpc and the halo concentration parameter $c=
7.2^{+5.0}_{-5.2}$. In the inner ICP regions we find $\bar{k}_c
= 0.21^{+0.48}_{-0.12}\times 10^{-2}$. Throughout the cluster
body we derive the entropy slope $a=0.71^{+0.05}_{-0.05}$. In
the outskirts we obtain $k_B T_R = 2.1^{+0.8}_{-0.6}$ keV and
$n_R = 2.0^{+0.2}_{-0.2}\times 10^{-5}$ cm$^{-3}$, yielding
$k_R = 2850^{+1370}_{-940}$ keV cm$^{2}$. Correspondingly, we
obtain $k_c= 6^{+18}_{-4}$ keV cm$^2$.

\subsection{Abell 1689}

A1689 is a rich cluster at $z\approx 0.183$ that has been
studied in various spectral regions; a joint analysis of
\textsl{HST/ACS} and \textsl{Chandra} measurements has been
carried out by Lemze et al. (2008) to derive the profiles of
the ICP density and temperature. The cluster is centered on a
cD galaxy, is roughly spherical and appears fairly relaxed (but
see Andersson \& Madejski 2004). The X-ray luminosity is
$L_X\approx 10^{45}$ erg s$^{-1}$ in the energy band $0.5-7$
keV, due to the hot ICP with $k_B T\approx 9.4$ keV (Ebeling et
al. 1996; Xue \& Wu 2002; Andersson \& Madejski 2004). Strong
and weak lensing observations (see Lemze et al. 2008 and
references therein) show the projected mass profile to
continuously flatten toward the center, with a steep outer
profile that require a high concentration $c\approx
13.7^{+1.4}_{-1.1}$  (see Broadhurst et al. 2008 and references
therein), the hallmark of an early transition.

With the SM we have fitted the ICP temperature profile observed
with \textsl{XMM-Newton} by Snowden et al. (2008), and the
surface brightness profile observed by Mohr et al. (1999) and
by Lemze et al. (2008). In agreement with the latter authors,
our fits (see Fig.~4) do not require an entropy floor; any
central flattening may affect only radii $r\la 2\times
10^{-3}\, R\approx 4$ kpc.

We adopt the virial radius $R = 2.1$ Mpc fixed at the value
given by Lemze et al. (2008), and from the SM we find the halo
concentration parameter $c= 13.6^{+4.3}_{-4.3}$. In the inner
ICP regions we find $\bar{k}_c\approx 2.4^{+0.8}_{-0.8}\times
10^{-2}$. Throughout the cluster body we derive the low
empirical value of the entropy slope $a=0.80^{+0.06}_{-0.06}$
(consistent with the value $0.82^{+0.02}_{-0.02}$ measured by
Lemze et al. 2008), that we relate after Eqs.~(3) and (4) to
the high concentration. In the outskirts we obtain $k_B T_R=
4.4^{+0.6}_{-0.6}$ keV and $n_R = 3.2^{+0.1}_{-0.1}\times
10^{-5}$ cm$^{-3}$, yielding $k_R \approx 4360^{+590}_{-590}$
keV cm$^{2}$. Correspondingly, we obtain $k_c= 105^{+49}_{-49}$
keV cm$^{2}$.

Finally, in Fig.~5 we show the mass profile derived from the SM
after Eq.~(8) on using the parameters obtained from our fits;
this turns out to be in good agreement with the total mass
profile obtained from the lensing data by Lemze et al. (2008).

\begin{figure*}
\center\epsscale{0.9} \plottwo{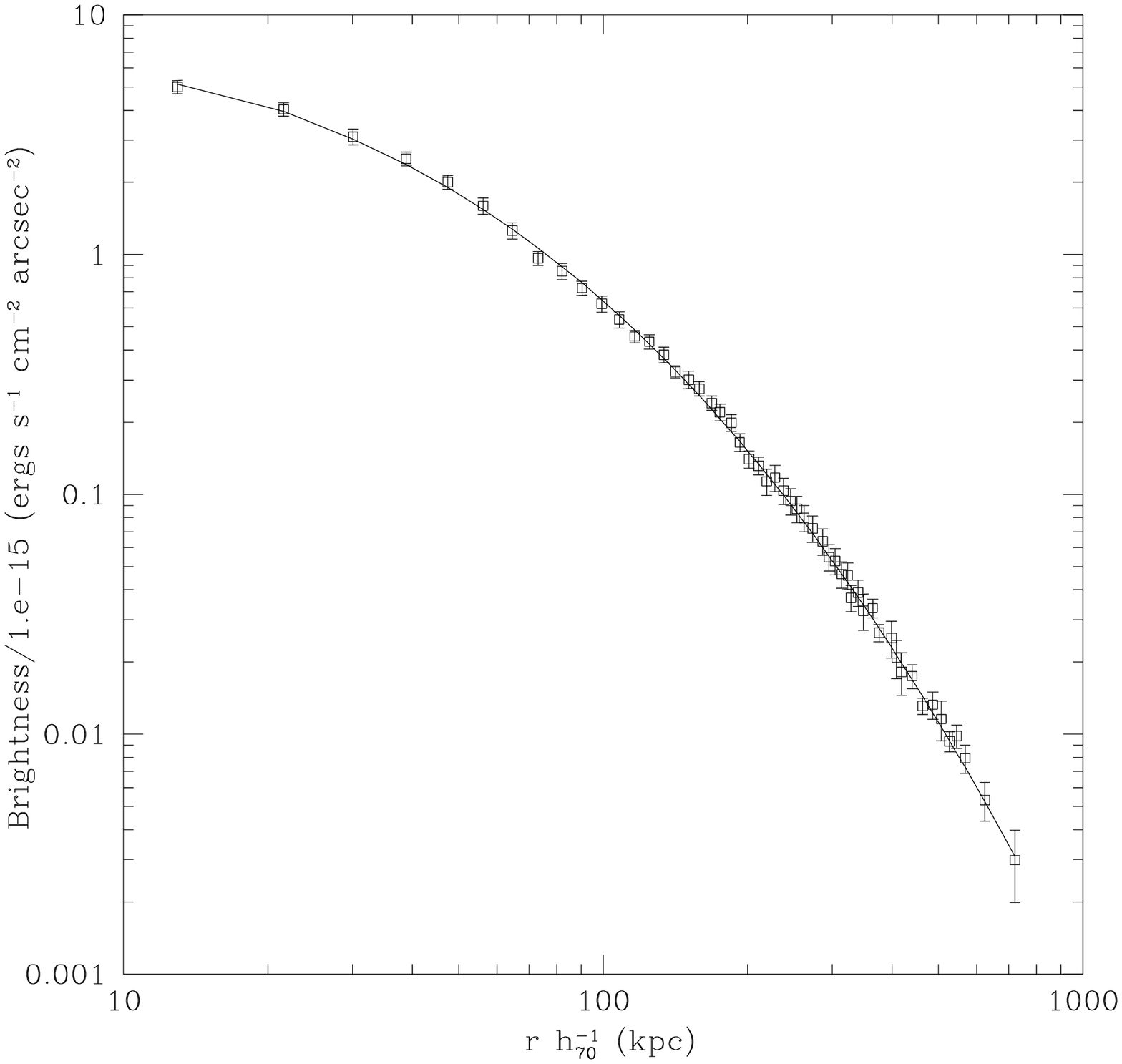}{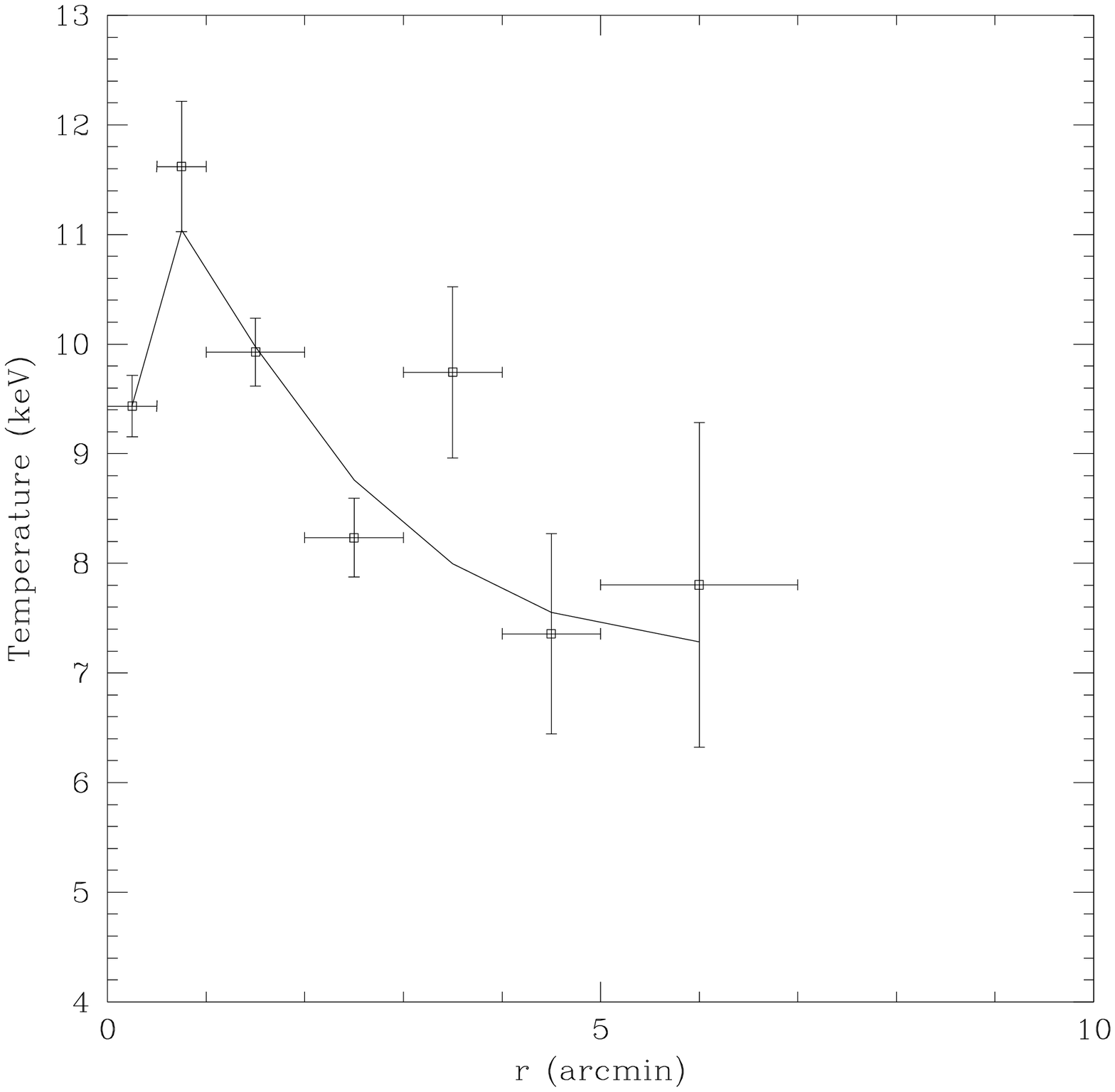}
\caption{A1689. {\it Left panel}: The solid line is our fit
($\chi^2$ = 19.8/49) to the brightness profile measured by
Lemze et al. (2008), on adopting the entropy profile given by
Eq.~(5). {\it Right panel}: The solid line is our fit ($\chi^2$
= 9.2/2) to the temperature profile measured by Snowden et al.
(2008), on adopting the entropy profile given by Eq.~(5).}
\end{figure*}

\begin{figure}
\center\epsscale{1.}\plotone{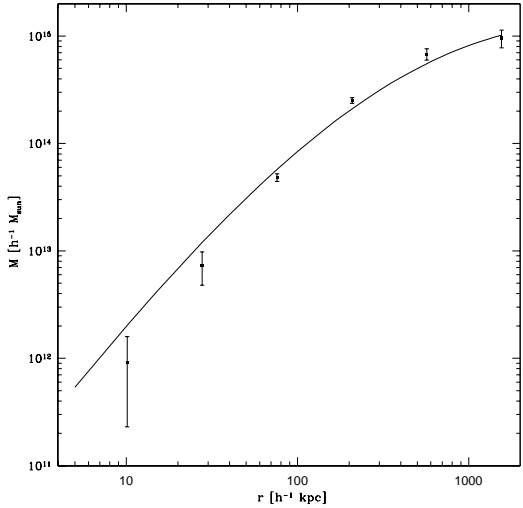} \caption{A1689. The solid
line is our fit to the mass profile measured by Lemze et al.
(2008) through gravitational lensing observations.}
\end{figure}

\subsection{Abell 1656 (Coma Cluster)}

A1656, the popular Coma Cluster, is a very rich cluster at
$z\approx 0.023$. Its ICP at temperature $k_B T\approx 8$ keV
is a powerful X-ray emitter with luminosity $L_X \sim 10^{45}$
erg s$^{-1}$ (see David et al. 1993) extending out to about
$1^{\circ}$ from the center (e.g., Briel et al. 2001).

Several determinations of the virial radius ranging between $2$
and $3$ Mpc are given in the literature (Castander et al. 2001;
{\L}okas \& Mamon 2003; Kubo et al. 2007; Gavazzi et al. 2009).
For our analysis, we have adopted the value of $2.2$ Mpc
reported by Gavazzi et al. (2009), and checked that our results
depend only weakly on this choice within $1$ standard
deviation.

From the SM we find the halo concentration parameter $c =
3.0^{+0.8}_{-0.8}$, the hallmark of a recent formation (see
\S~1). In the inner ICP regions we find $\bar{k}_c\approx
10^{+1}_{-1}\times 10^{-2}$. Throughout the cluster body we
derive the entropy slope $a = 1.30^{+0.46}_{-0.23}$. In the
outskirts we obtain $k_B T_R= 5.7^{+1.0}_{-1.0}$ keV and $n_R =
3.66^{+0.15}_{-0.15}\times 10^{-5}$ cm$^{-3}$, yielding $k_R =
5170^{+1060}_{-1045}$ keV cm$^{2}$. Correspondingly, we find
$k_c=520^{+160}_{-160}$ keV cm$^2$.

Our fits to the the brightness data observed with
\textsl{ROSAT} by Mohr et al. (1999) are obtained on using the
two entropy profiles given by Eqs.~(5) or by Eqs.~(6) and (7),
and are illustrated in Fig.~6; the $\chi^2$ values in the
caption strongly indicate the presence of an entropy floor with
extension $r_f = 250^{+44}_{-74}$ kpc. On the other hand, the
fit to the emission-weighted temperature profile (see Fig.~7)
obtained from \textsl{XMM-Newton} observations by Snowden et
al. (2008) is roughly isothermal, and within its resolution
does not require by itself an entropy floor.

All the above concurs to a picture of A1656 as a halo that has
just collapsed and undergone major mergers, as confirmed by the
structured features around the main galaxies NGC4874 and
NGC4889 and by the ongoing fall of the NGC4839 galaxy group
into the main body of the cluster (Adami et al. 2005). A1656
also exhibits extended radio emissions, including a young radio
halo close to center and an outer radio relic located in the SW
direction beyond the NGC4839 galaxy group (Giovannini et al.
1991, and references therein).

From Eq.~(8) we derive an overall mass $M_R=
1.24^{+0.44}_{-0.66}\times 10^{15}\,M_{\odot}$, in agreement
with the value of $9.7^{+6.1}_{-3.5}\times 10^{14}\, M_{\odot}$
obtained by Gavazzi et al. (2009) from gravitational lensing.

\begin{figure*}
\center\epsscale{0.9} \plottwo{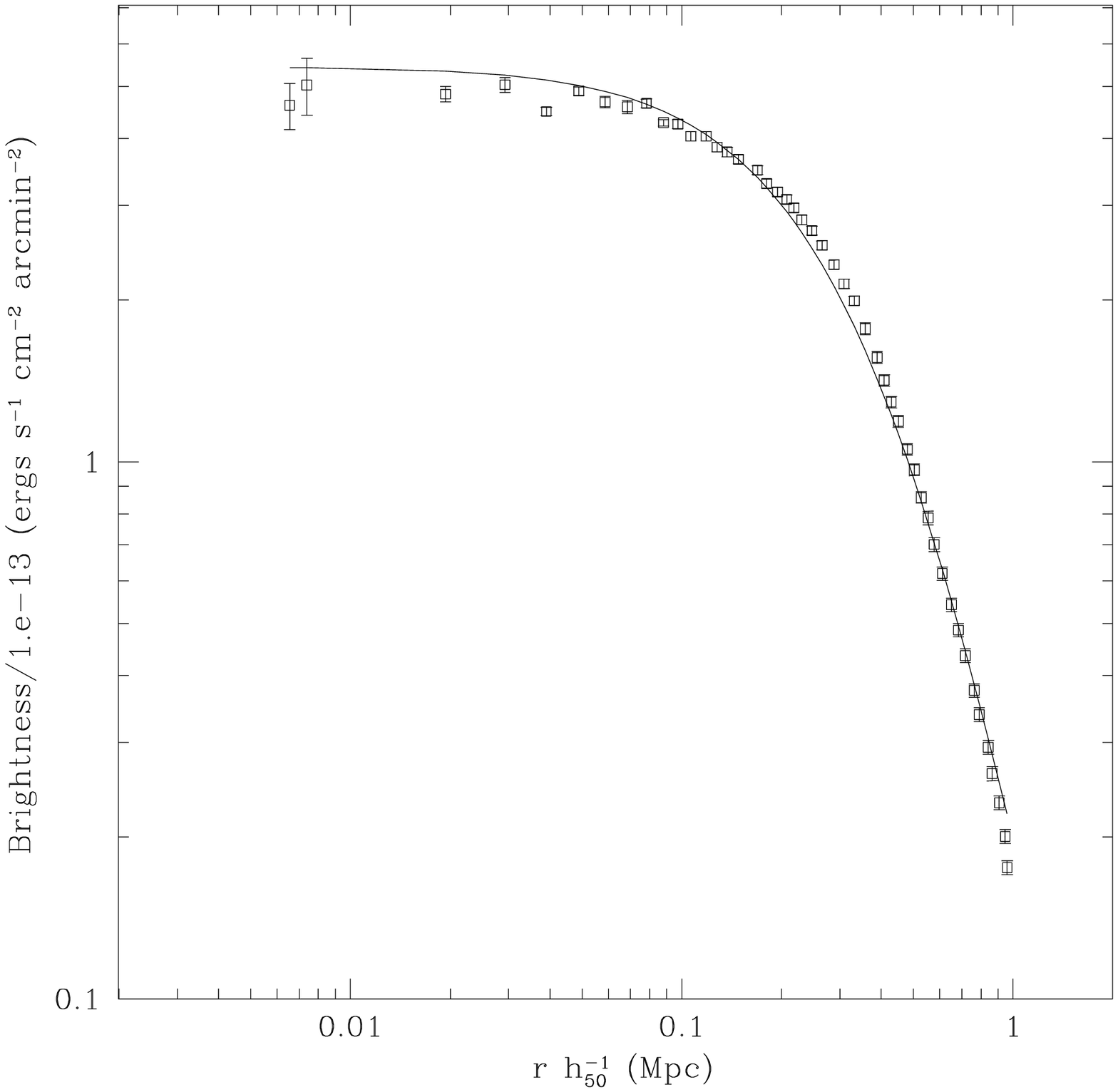}{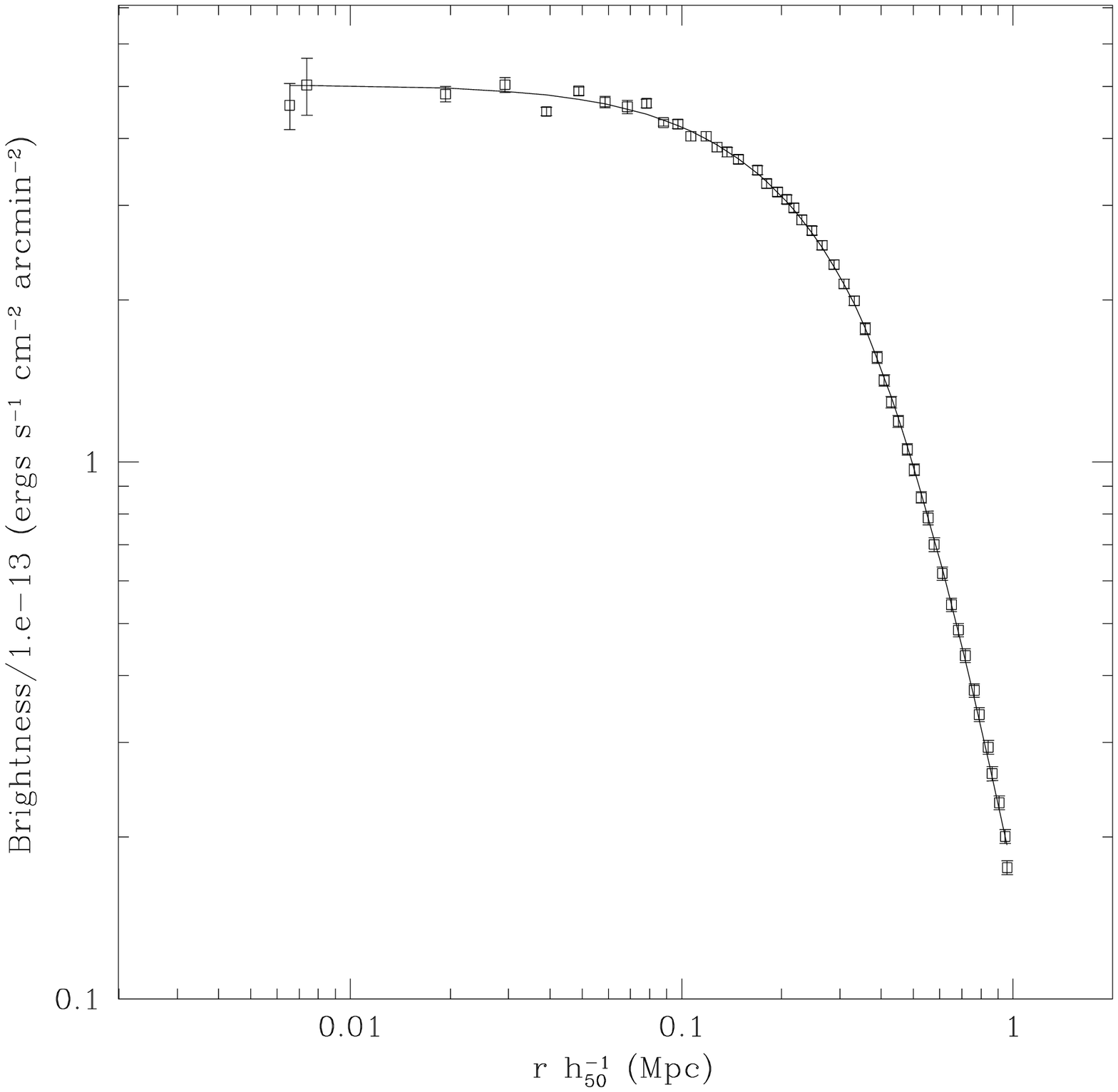}
\caption{A1656. {\it Left panel}: The solid line is our fit
($\chi^2$ = 408.5/44) to the brightness profile measured by
Mohr et al. (1999), on adopting the entropy profile given by
Eq.~(5). {\it Right panel}: The solid line is our fit ($\chi^2$
= 53.0/43) on adopting the entropy profile given by Eqs.~(6)
and (7).}
\end{figure*}

\begin{figure*}
\center\epsscale{0.9} \plottwo{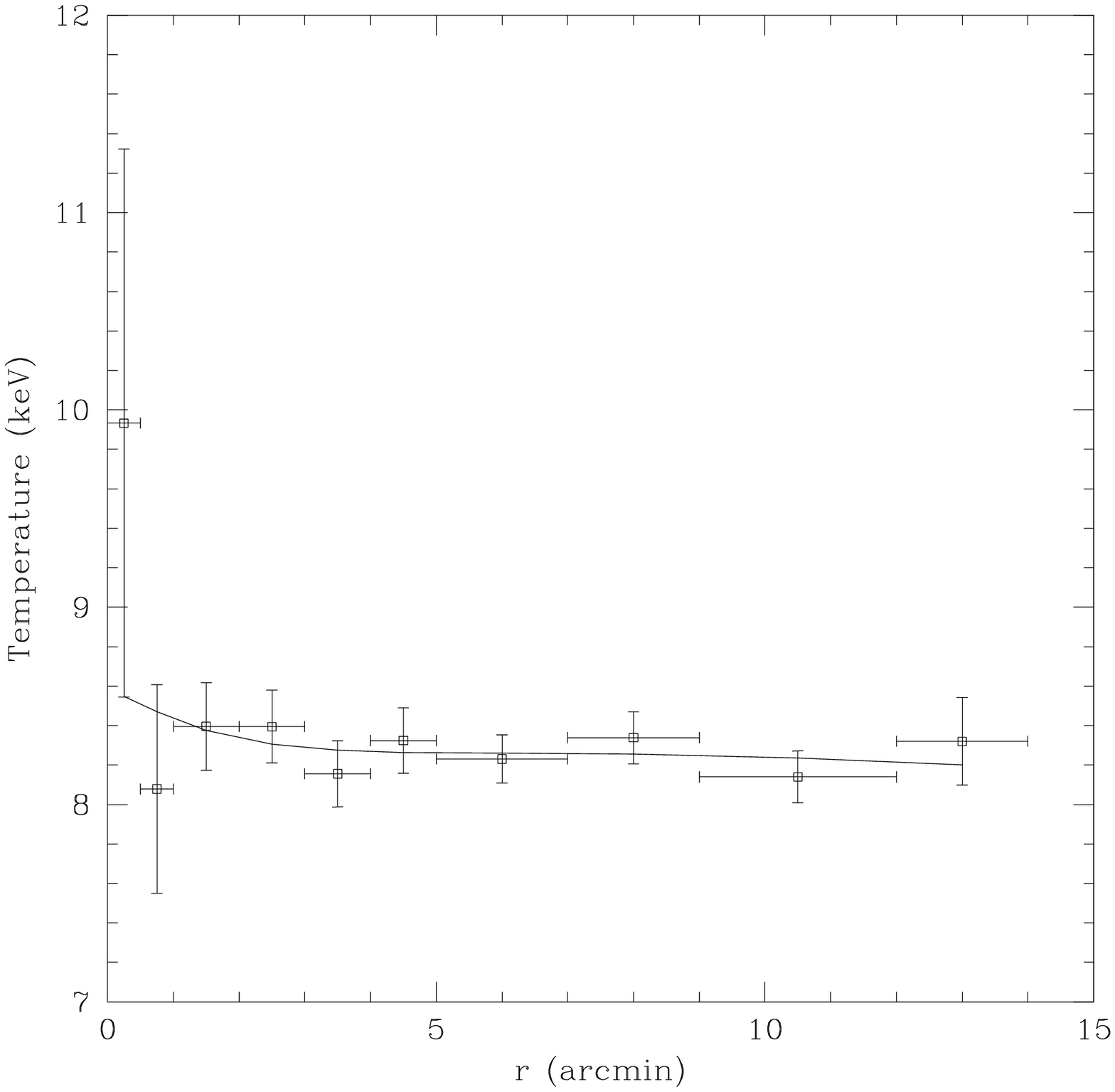}{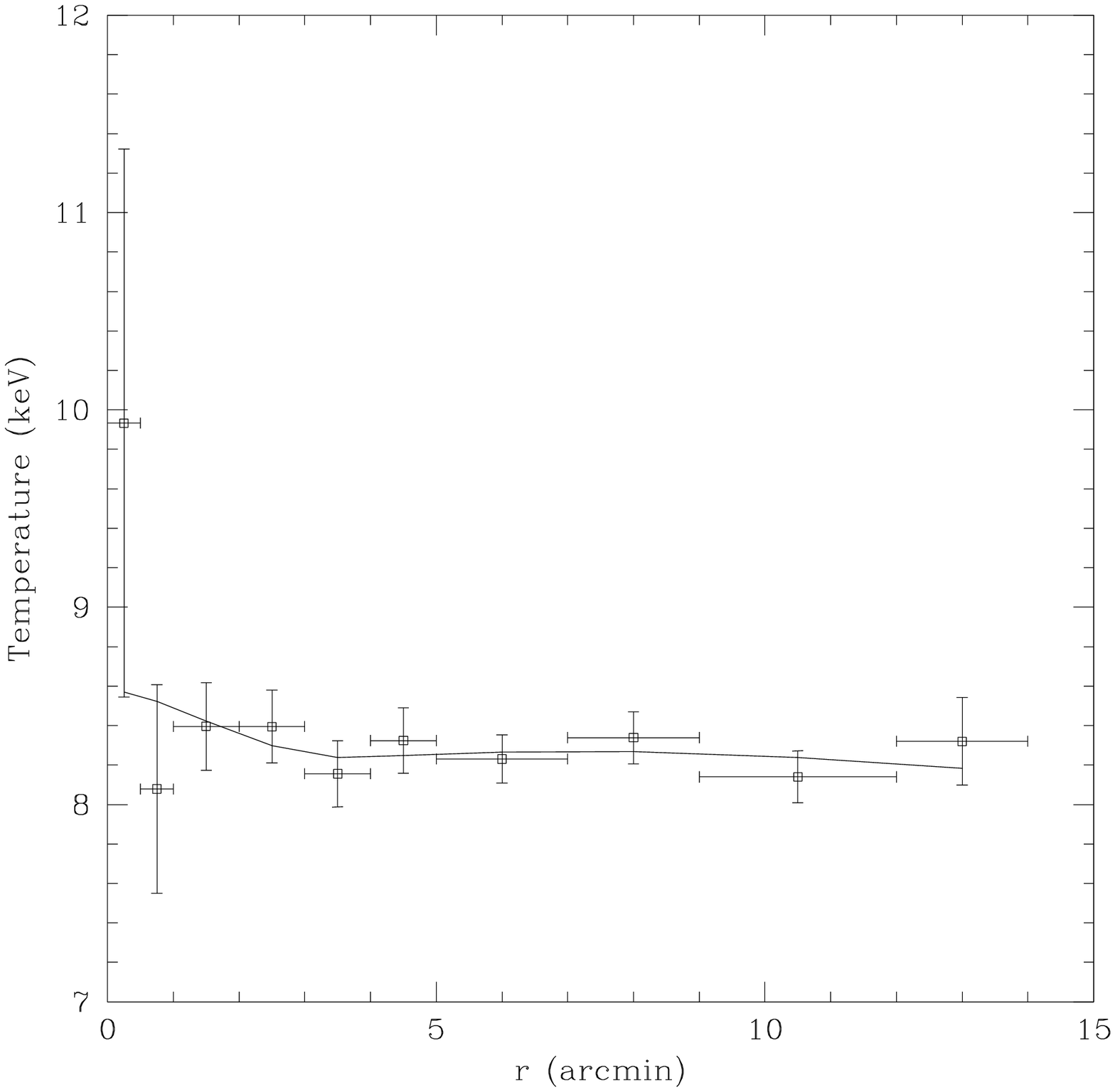}
\caption{A1656. {\it Left panel}: The solid line is our fit
($\chi^2$ =  3.7/6) to the temperature profile measured by
Snowden et al. (2008), on adopting the entropy profile given by
Eq.~(5). {\it Right panel}: The solid line is our fit ($\chi^2$
= 3.7/5) on adopting the entropy profile given by Eqs.~(6) and
(7).}
\end{figure*}

\subsection{Abell 2256}

A2256 at $z\approx 0.06$ is a \emph{complex} cluster, as
highlighted by several studies in various spectral bands; it is
a strong X-ray emitter with luminosity $L_X \approx 10^{45}$
erg s$^{-1}$. We have used the SM to fit the \textsl{ROSAT}
brightness distribution by Mohr et al. (1999) and the
temperature profile observed with \textsl{XMM-Newton} by
Snowden et al. (2008), on using the entropy profiles given by
Eqs.~(5) or by Eqs.~(6) and (7); our results are illustrated in
Fig.~8 and 9.

From the SM, we find the virial radius $R = 2.2^{+0.3}_{-0.3}$
Mpc, and the halo concentration parameter $c=2.7^{+1.7}$. In
the inner ICP regions we find
$\bar{k}_c=6.2^{+3.9}_{-3.1}\times 10^{-2}$. Throughout the
cluster body we derive the entropy slope $a =
1.48^{+0.35}_{-0.29}$. In the outskirts we obtain $k_B T_R =
4.4^{+0.9}_{-0.9}$ keV and $n_R=3.64^{+0.04}_{-0.22}\times
10^{-5}$ cm$^{-3}$, yielding $k_R=4000^{+1097}_{-980}$ keV
cm$^{2}$. Correspondingly, we find $k_c=248^{+224}_{-185}$ keV
cm$^2$.

Both the brightness and temperature distributions call for an
entropy floor; basing on Eqs.~(6) and (7) we derive a floor
radius $r_f=264^{+102}_{-80}$ kpc from the brightness, an
evidence reinforced by the value $r_{f}=265^{+80}_{-170}$ kpc
we obtain from the temperature profile. Its introduction allows
the SM to fit well the structured temperature profile of A2256;
this features a temperature decrement similar to a CC cluster
(e.g., Piffaretti et al. 2005; Leccardi \& Molendi 2009), but
at small radii $T(r)$ reverses its trend and increases toward
the center. Such a behavior is understood from the relation
$T(r)\propto k(r)\, n(r)^{2/3}$; in the inner region where the
entropy is constant, the temperature is expected to decrease
outwards following the density, while for $r>r_f$ the entropy
starts to increase and to dominate the density decrement, so
raising the temperature out to a peak at $r\approx 350$ kpc
(about $5'$); beyond the peak the density steepens and offsets
the entropy rise. The central temperature behavior suggests
that the energy delivered by a merger has remolded the whole
inner structure, and hence that the ICP is itself
thermodynamically young within a dynamically young DM halo.

The halo's young age is supported by the low values $c\approx
4$ of the DM concentration, as expected in clusters with a
recent transition from fast collapse to slow accretion (see
\S~1). It is also indicated by the recent and intense merger
activity that characterize A2256, as pinned down by
\textsl{ROSAT}, \textsl{Chandra} and \textsl{XMM-Newton}
observations (Briel et al. 1991; Sun et al. 2002; Berrington et
al. 2002; Miller et al. 2003; Bourdin \& Mazzotta 2008). Such a
merger activity is quite recent as indicated by the presence of
relativistic electrons with lifetime of $10^{-1}$ Gyr that
enlighten the radio halo and the very extended, bright relic in
the NW region of the cluster (Bridle \& Formalont 1976; Kim
1999; Clarke \& Ensslin 2006). All that concurs with the high
central entropy level pinned down by the SM analysis.

\begin{figure*}
\center\epsscale{0.9} \plottwo{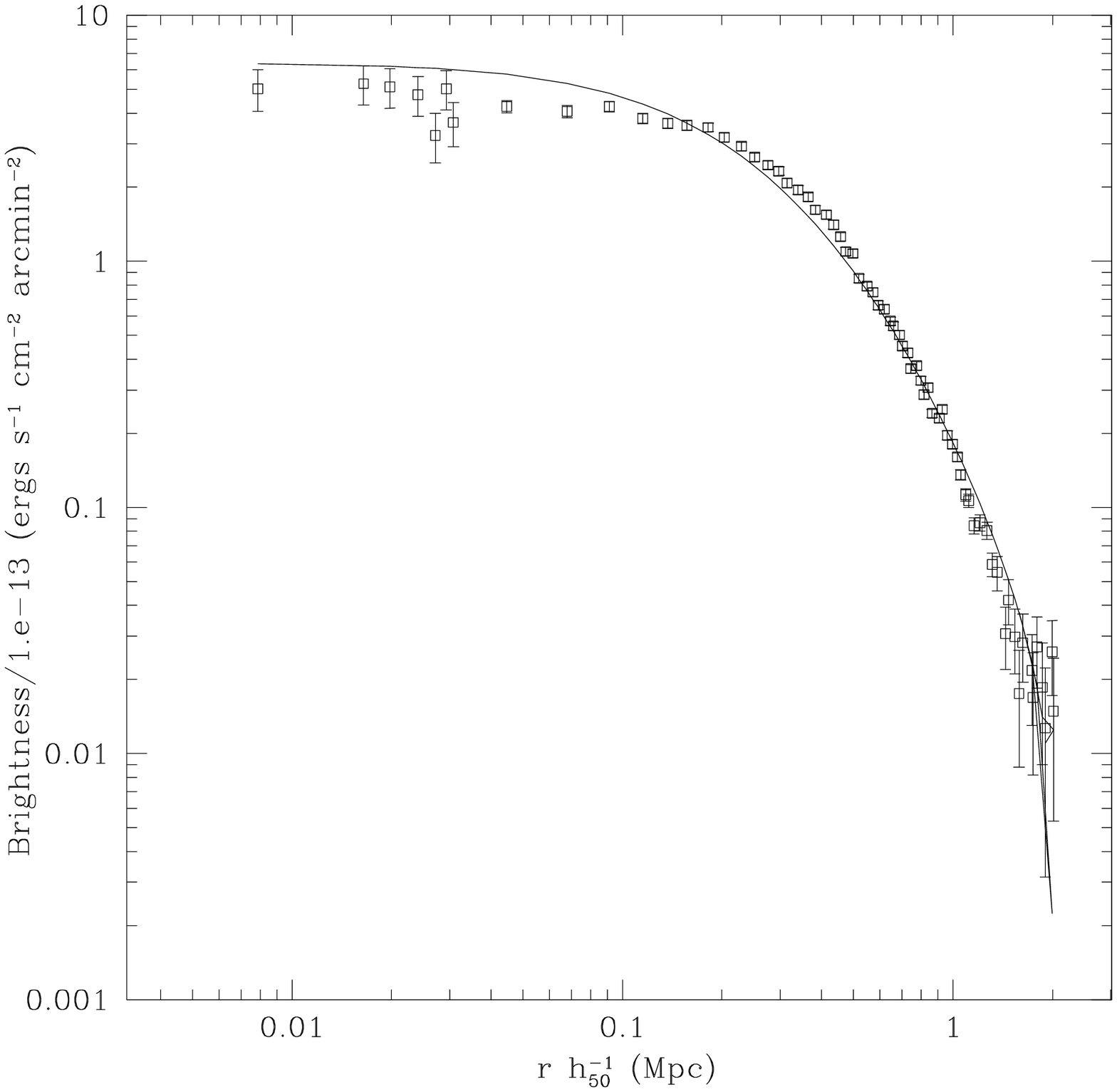}{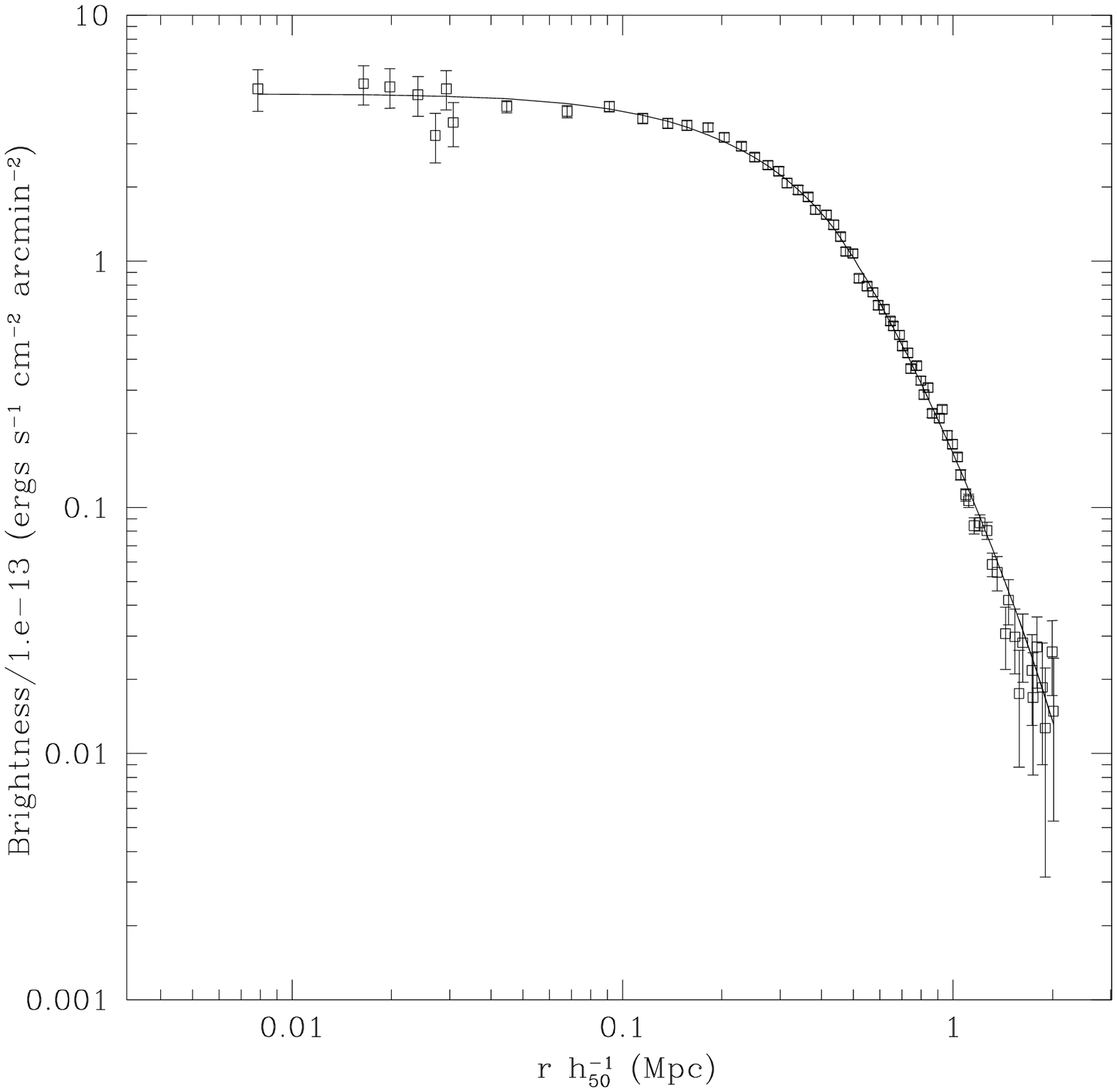} \caption{
A2256. {\it Left panel}: The solid line is our fit ($\chi^2$ =
434.1/65) to the brightness distribution measured by Mohr et
al. (1999), on adopting the entropy profile given by Eq.~(5).
{\it Right panel}: The solid line is our fit ($\chi^2$ =
109.7/64) on adopting the entropy profile given by Eqs.~(6) and
(7).}
\end{figure*}

\begin{figure*}
\center\epsscale{0.9} \plottwo{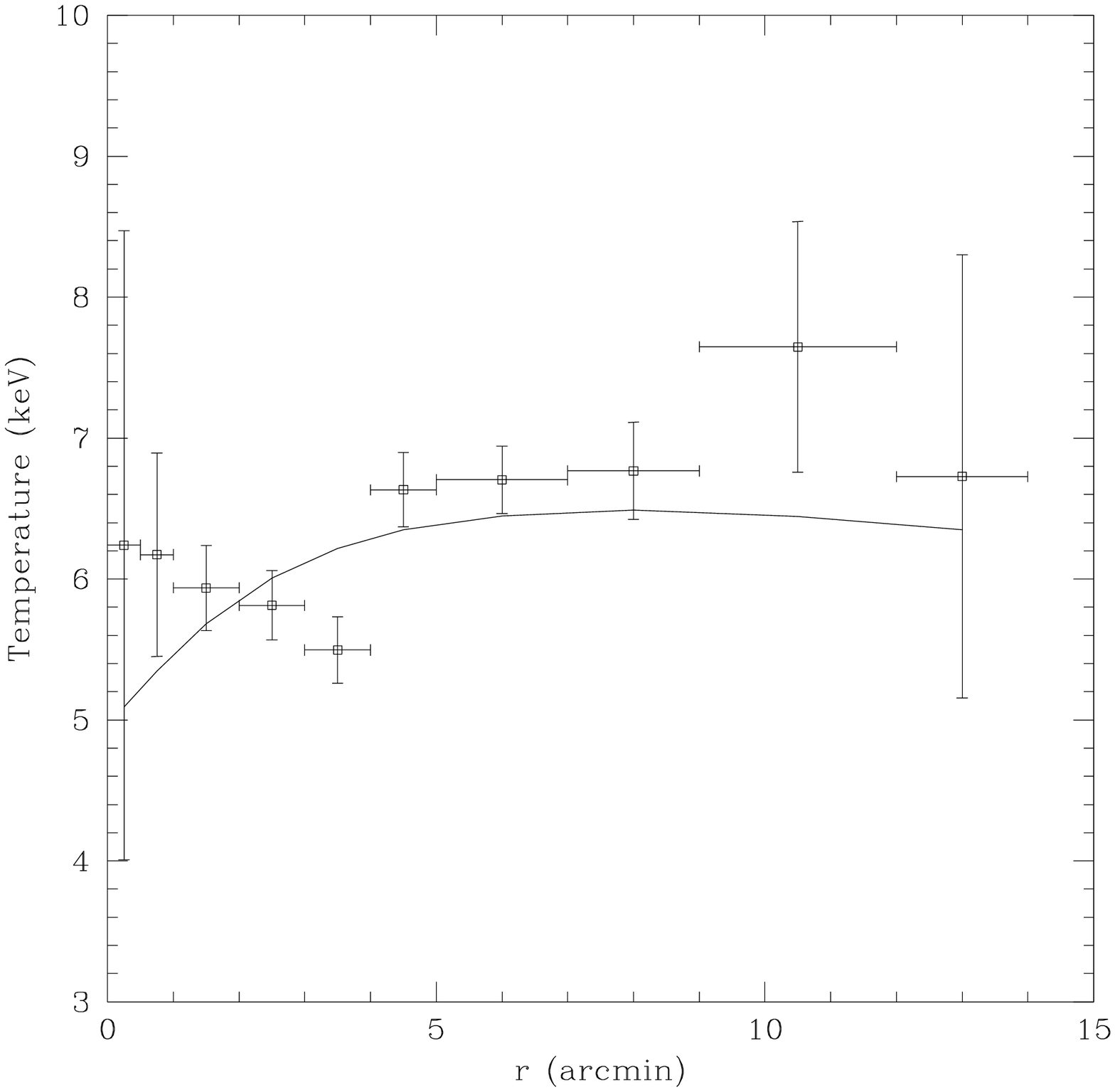}{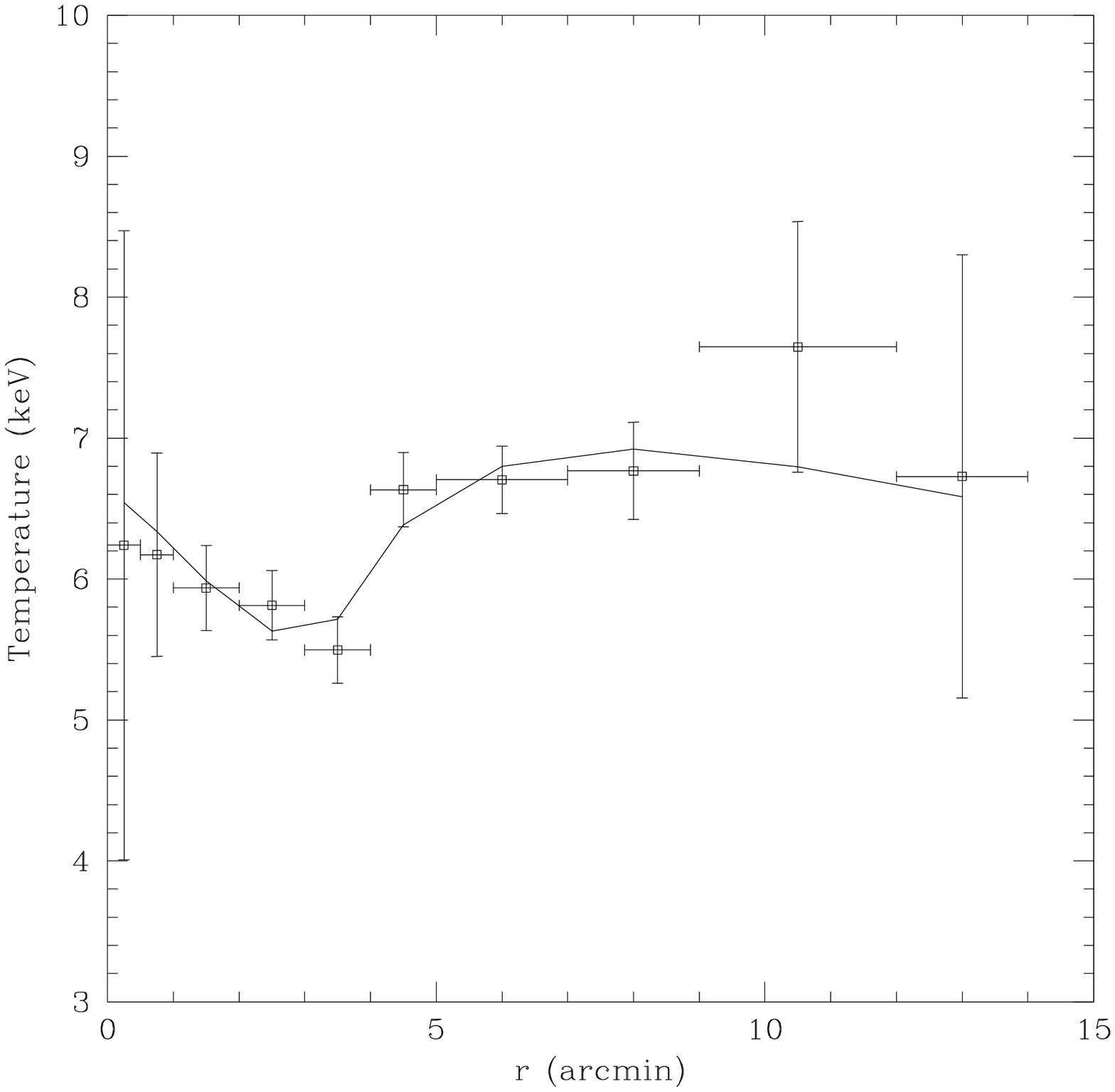}
\caption{A2256. {\it Left panel}: The solid line is our fit
($\chi^2$ = 12.4/5) to the temperature profile measured by
Snowden et al. (2008), on adopting the entropy profile given by
Eq.~(5). {\it Right panel}: The solid line is our fit ($\chi^2$
= 3.6/4) on adopting the entropy profile given by Eqs.~(6) and
(7). Note that the first bin covers the range out to about
$0.05\,h_{50}^{-1}$ Mpc in Fig.~8.}
\end{figure*}

\subsection{Abell 644}

A644 is a radio quiet cluster at $z\approx 0.07$. The X-ray
imaging with \textsl{Chandra} by Buote et al. (2005) shows
that, while the outer ICP is smooth and relaxed, the central
$\sim 10^2$ kpc features \emph{very complex} and interesting
conditions. Although the latter do not bear strong marks of
past radio activity like X-ray cavities or filaments, the peak
of the X-ray emission is found to be shifted by about $60$ kpc
from the cluster centroid (see also Bauer \& Sarazin 2000),
beyond the cD galaxy located at about $40$ kpc. The temperature
profiles differ when centered on the centroid or on the X-ray
peak; the former profiles deviates only weakly from a constant,
while the latter shows a behavior similar to, if sharper than
A2256, in that it decreases considerably toward the center
before reversing its course and rising again at small radii
(Buote et al. 2005). Our SM fits to this temperature profile,
obtained on using Eq.~(5) or Eqs.~(6) and (7), are illustrated
in Fig.~10; the fit to the brightness profile centered on the
cluster centroid is illustrated in Fig.~11.

We find the virial radius $R= 2.1^{+0.4}_{-0.4}$ Mpc and the
halo concentration parameter $c=3.9_{-0.2}$. At the center we
find $\bar{k}_c=0.7^{+0.1}_{-0.1}\times 10^{-2}$. On using
Eqs.~(6) and (7) we derive a radius $r_f=61^{+36}_{-41}$ kpc,
whose need is substantiated by the $F$-test at $99.9\%$ level
and also strengthened by the fit to the \textsl{Chandra}
brightness profile centered on the X-ray peak (D. Buote,
private communication) that consistently yields
$r_f=66^{+8}_{-9}$ kpc. Throughout the cluster body we derive
the entropy slope $a=1.06^{+0.11}_{-0.11}$. In the outskirts we
find $k_B T_R=4.9^{+1.8}_{-1.8}$ keV and $n_R=
2.88^{+0.19}_{-0.19}\times 10^{-5}$ cm$^{-3}$, yielding $k_R =
5200^{+2140}_{-2140}$ keV cm$^{2}$. Correspondingly, we find
$k_c=36^{+20}_{-20}$ keV cm$^{2}$. Finally, from Eq.~(8) we
derive the total mass $M= 1.2^{+0.7}_{-0.7}\times
10^{15}\,M_{\odot}$.

Note that if one insisted on applying the SM also to
temperature and brightness profiles from the centroid on the
basis of Eqs.~(6) and (7), one would obtain $r_f=104^{+4}_{-4}$
kpc and a related lower bound $k_c\approx 124^{+120}$ keV
cm$^2$. The large variance in the above parameters entering the
entropy floors signals complex substructures, that may be
interpreted as a high density, low entropy clump (`cold drop')
around the X-ray peak. Our results agrees with the analysis by
Buote et al. (2005), who in terms of two differently centered
$\beta$-models (Cavaliere \& Fusco-Femiano 1976) find two core
sizes similar to our extensions $r_f$.

In view of the lack of radio emission and X-ray cavities, this
complexity may be understood in terms of a merger having just
remolded an inner region of the ICP (see Henning et al. 2009).
We stress that such ICP substructures constitute a common, but
progressively more pronounced trait, of A1656, A2256 and A644.

\begin{figure*}
\center\epsscale{0.9} \plottwo{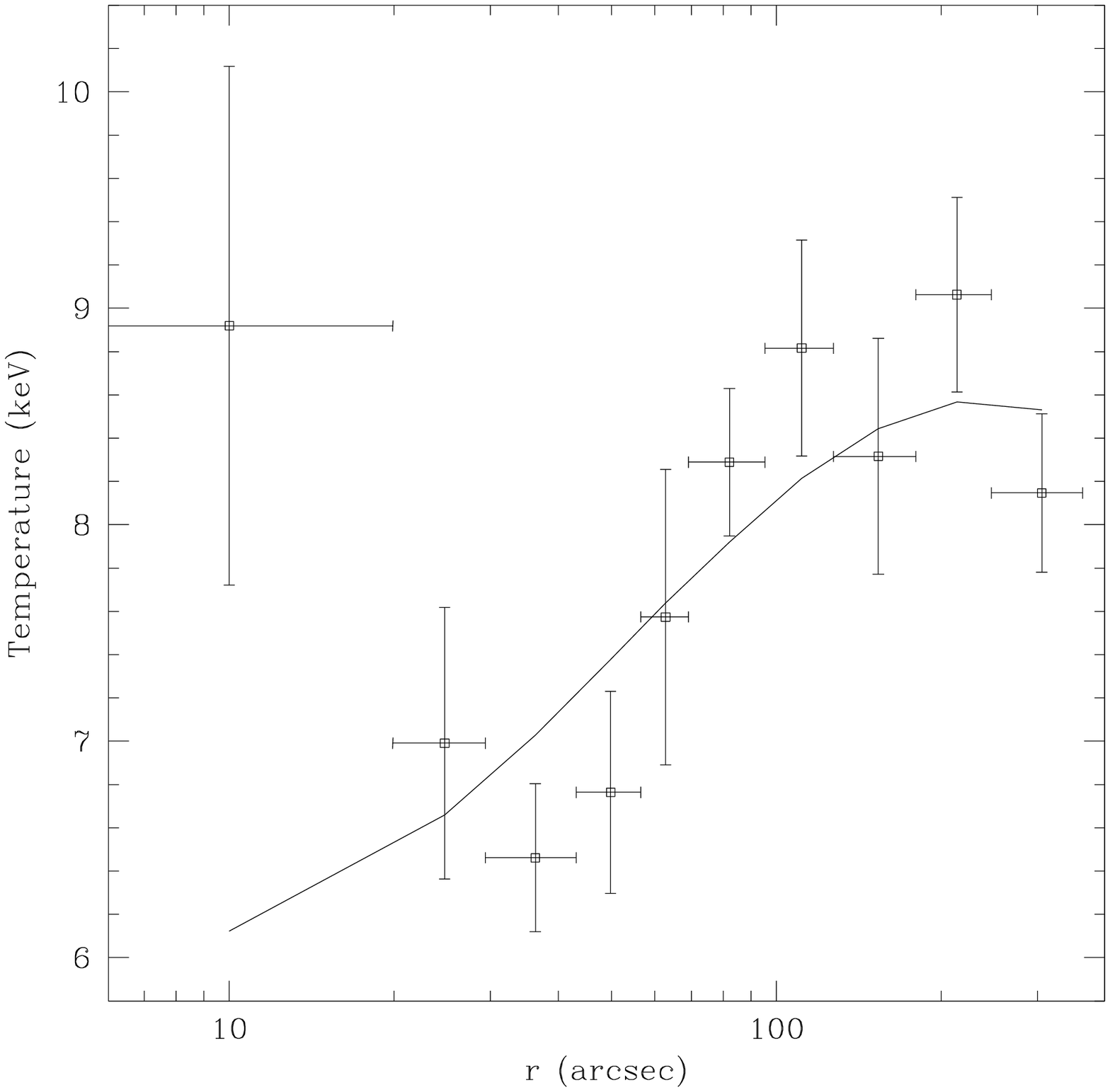}{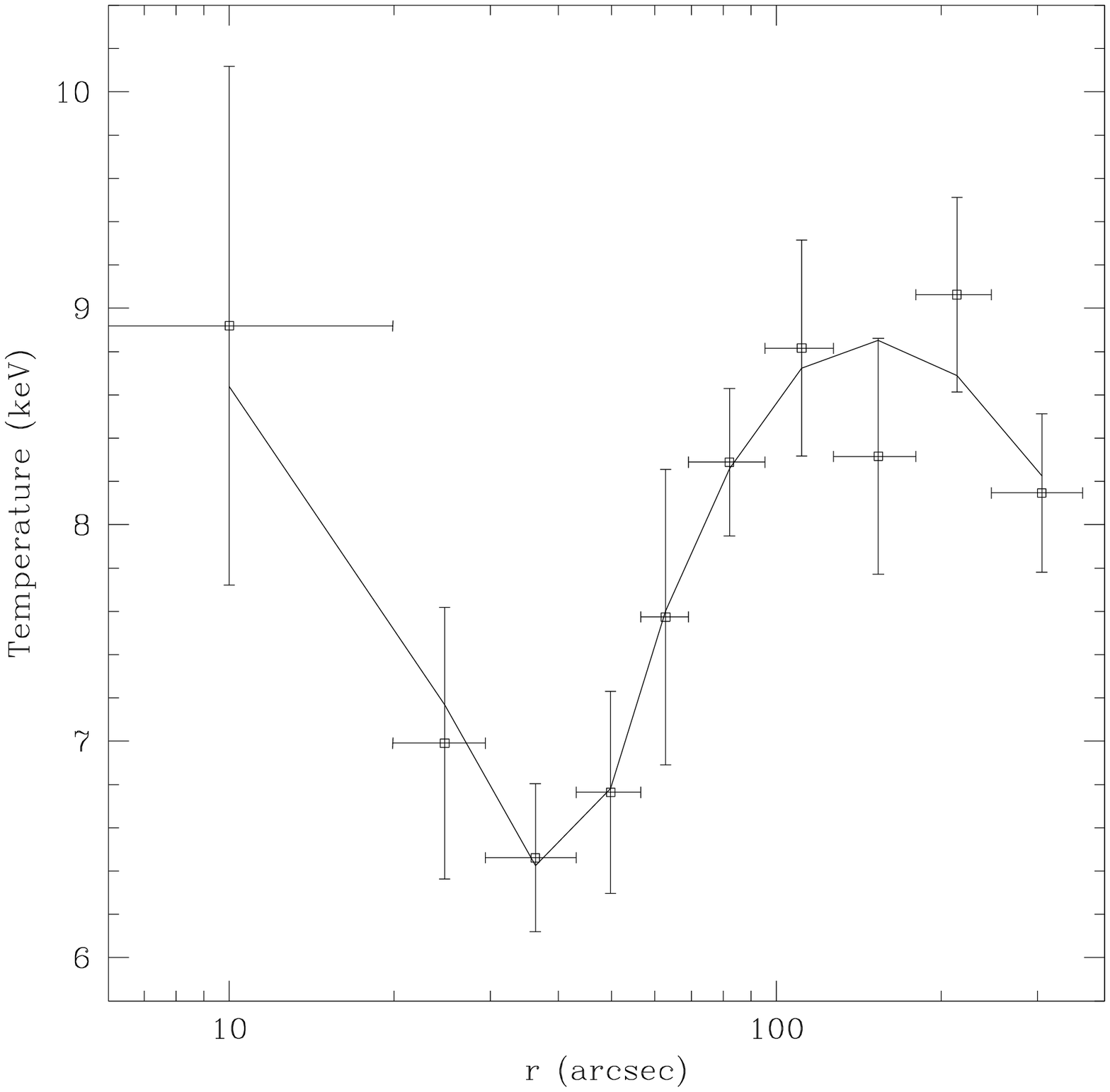}
\caption{A644. {\it Left panel}: The solid line is our fit
($\chi^2$ = 15.2/5) to the temperature profile centered on the
X-ray peak measured by Buote et al. (2005), on adopting the
entropy profile given by Eq.~(5). {\it Right panel}: The solid
line is our fit ($\chi^2$ = 1.97/4) on adopting the entropy
profile given by Eqs.~(6) and (7). Note that the first bin
covers the range out to about $0.04\,h_{50}^{-1}$ Mpc in
Fig.~11.}
\end{figure*}

\begin{figure*}
\center\epsscale{0.9} \plottwo{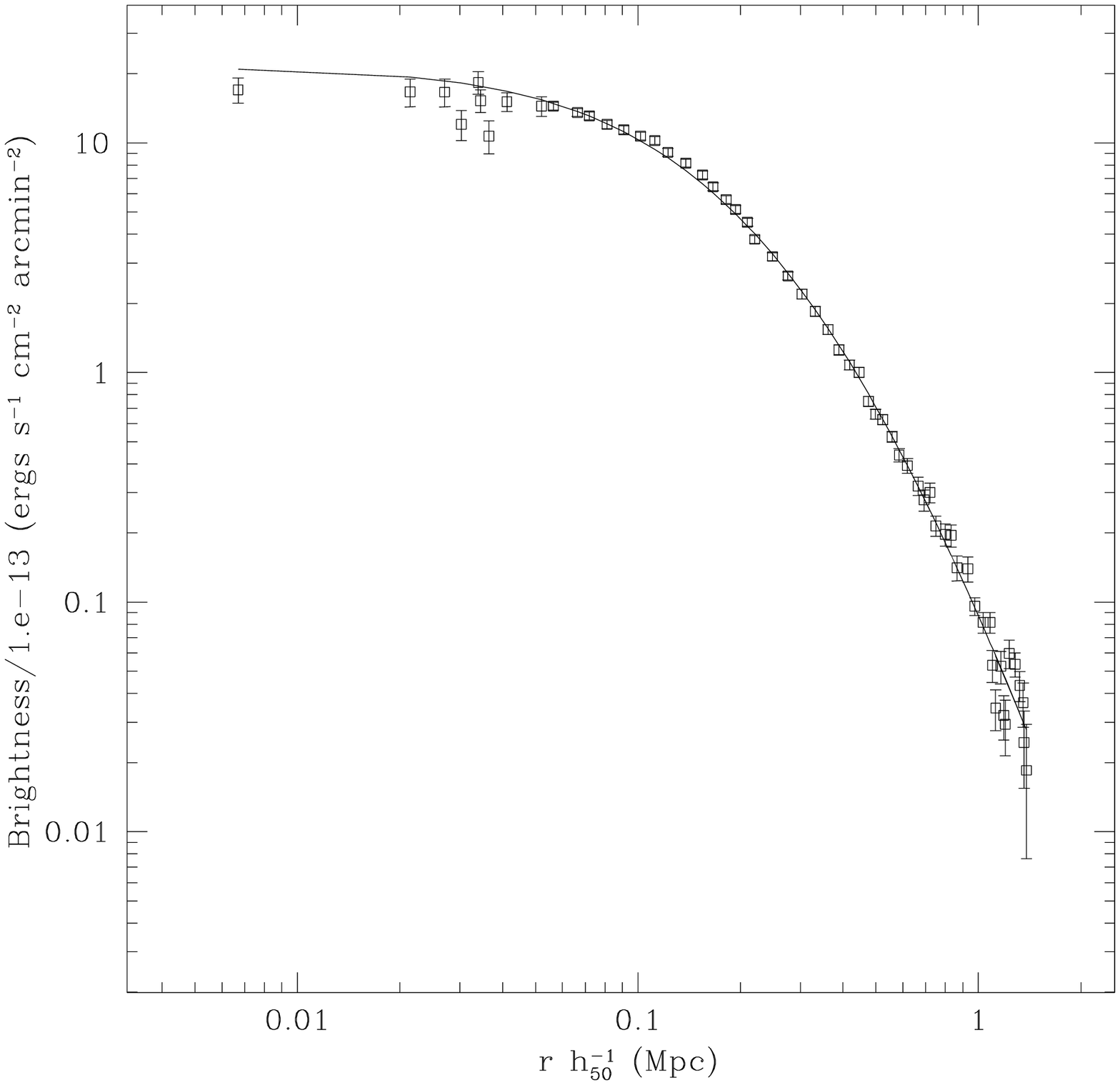}{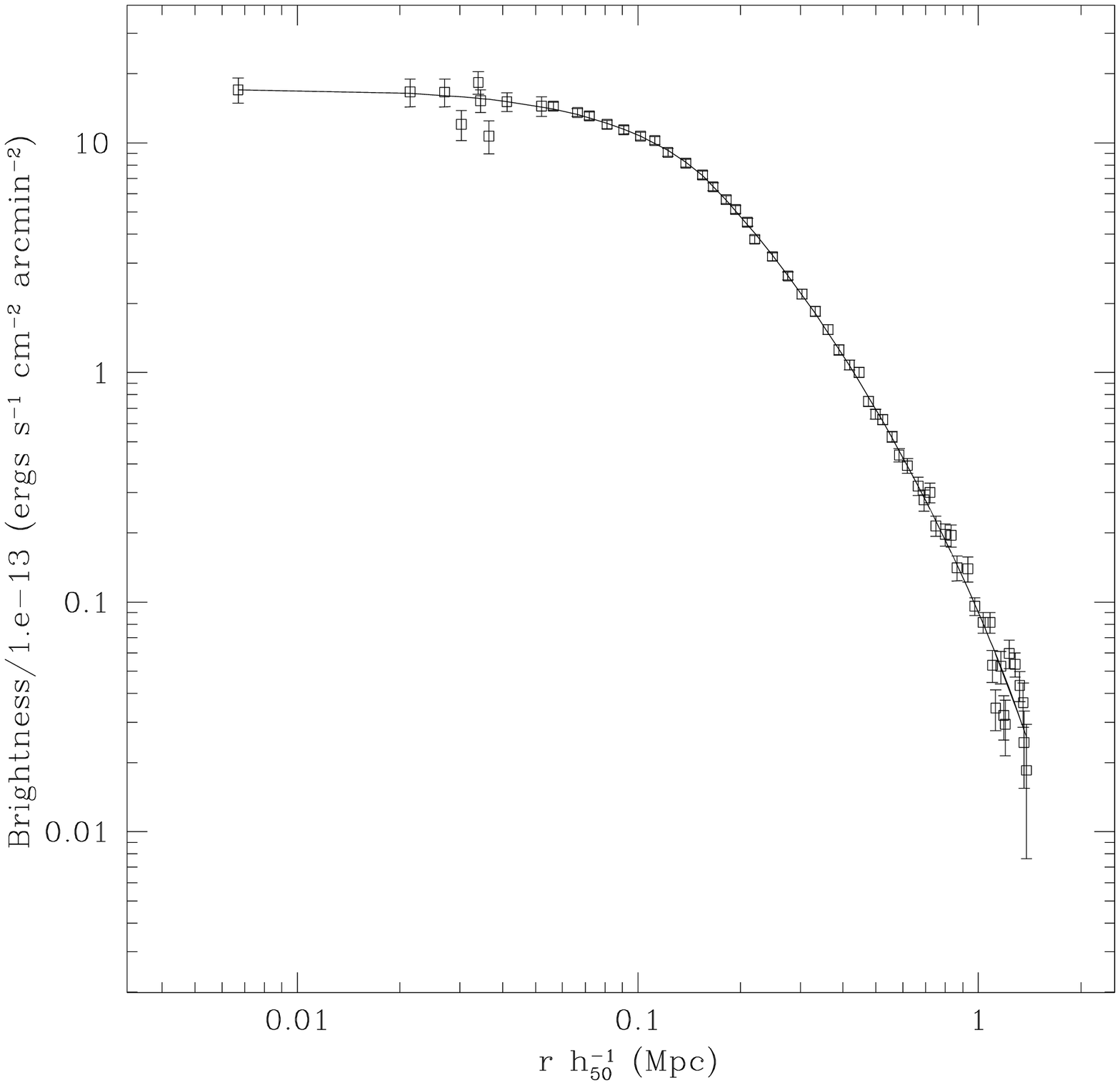}
\caption{A644. {\it Left panel}: The solid line is our fit
($\chi^2$ = 120.7/56) to the brightness profile measured by
Mohr et al. (1999), on adopting the entropy profile given by
Eq.~(5). {\it Right panel}: The solid line is our fit ($\chi^2$
= 76.3/55) on adopting the entropy profile given by Eq.~(6) and
(7).}
\end{figure*}

\section{Discussion and Conclusions}

In this paper we have analyzed with the Supermodel (SM) the
profiles of X-ray temperature and surface brightness of the
IntraCluster Plasma (ICP) in a set of six clusters (adding to
the three ones preliminarily reported in CLFF09) with existing
detailed data. We have shown how effective is our SM to
represent and understand the main Cool Core /Non Cool Core
\emph{dichotomy} in terms of two physical parameters marking
the full ICP entropy profile: the central value $k_c$, and the
outer slope $a$ (see Figs.~1-7). Moreover, the SM makes sense
of more \emph{structured} profiles (see Figs.~8-11) in terms of
the additional, physical parameter $r_f$ marking the extension
of the entropy floor.

The working of the SM may be reduced to the bones as follows.
The spatial scale for the temperature peak in the CCs, and for
the outward temperature and density declines in all clusters,
is set by the underlying \emph{gravitationally} dominant DM
distribution; specifically, such a scale is set by the peak of
the DM velocity dispersion $\sigma^2(r)$ at the position
$r_m\approx r_{-2}$, that divides the `inner' from the `outer'
cluster regions (see Figs.~1 and 3 of CLFF09). In fact, the SM
ensures that the approximation $T\simeq\sigma^2/\beta_m$ [with
$\beta_m\equiv \mu m_p\sigma^2(r_m)/k_BT(r_m)$, see CLFF09] is
to hold closely for all clusters around $r_m$, and over a wide
radial range for the CCs\footnote{The formal reason why around
$r_m$ the temperature $T(r)$ mirrors so well the DM dispersion
$\sigma^2(r)$ is that the latter can be expressed quite
similarly to Eq.~(2), except for the constant term within
square brackets that is anyway negligible in cluster bodies
(see CLFF09), and particularly so for the CCs.}. In the inner
ICP regions, the temperature and density profiles are governed
primarily by the central entropy level $k_c$ set by
\emph{thermodynamical} events, i.e., energy discharged and
blasts driven by major mergers or AGN outbursts; in a number of
cases the latter imprint substructure on the small scale $r_f$
comparable to $r_m$. A novel feature of the SM (relative to
handy isothermal or polytropic $\beta$-models where the entropy
is assumed to be a functional $k\propto n^{\Gamma-5/3}$ of the
density, see Cavaliere \& Fusco-Femiano 1976) is constituted by
the \emph{articulated} radial entropy run entering Eq.~(2).

We have collected in Table~1 the SM fitting parameters for all
clusters in our set. By inspection it is apparent a correlation
between the inner ICP profile type (marked by the CC/NCC tags)
with the central entropy level $k_c$, the outer entropy slope
$a$, and the DM concentration $c$; high values of $c$ and low
values of $a$ and $k_c$ correspond to the CC class, while the
opposite trend holds for NCC. We understand these trends in the
framework of two-stage cluster formation (see \S~1) as follows.
For example, low values of $a$ correspond to high values of
$b_R$ (see Eq.~4) owing to low values of $\Delta\phi$ (see
Eq.~3); these are related to high concentrations $c=
3.5\,(1+z_t)$ (see \S~1 and 2), which imply early transition
redshifts $z_t$, i.e., an old age. We stress that in the six
clusters considered here the outer parameters $a$ and $c$ turn
out to be related to the inner parameter $k_c$ (and $r_f$ when
applicable) in terms of the cluster age, as explained below.

\begin{deluxetable*}{lllcccccccccccccc}
\tabletypesize{}\tablecaption{Fitting parameters from the SM
analysis} \tablewidth{0pt} \tablehead{\colhead{Cluster} & &
\colhead{Class} & &\colhead{$c$} & & \colhead{$a$} & &
\colhead{$\bar{k}_c$ $[10^{-2}]$} & & \colhead{$r_f$ [kpc]} & &
\colhead{$R$ [Mpc]} & & \colhead{$k_B T_R$ [keV]}}\startdata
A2199 && CC        && $6.7^{+1.0}_{-1.0}$  &&
$0.95_{-0.01}^{+0.01}$  && $0.39^{+0.16}_{-0.16}$  && $\la 2$
&& $2.1_{-0.4}$
&& $1.93^{+0.05}_{-0.05}$ \\
A2597 && CC        && $7.2^{+5.0}_{-5.2}$  &&
$0.71^{+0.05}_{-0.05}$  && $0.21^{+0.48}_{-0.12}$  &&
$50^{+7}_{-7}$      && $1.9^{+0.4}_{-0.4}$  &&
$2.1^{+0.8}_{-0.6}$    \\
A1689 && CC        && $13.6^{+4.3}_{-4.3}$ &&
$0.80^{+0.06}_{-0.06}$  && $2.4^{+0.8}_{-0.8}$     && $\la 4$
&& $2.1$ && $4.4^{+0.6}_{-0.6}$    \\
A1656 && NCC       && $3.0^{+0.8}_{-0.8}$  &&
$1.30^{+0.50}_{-0.24}$     && $10^{+1}_{-1}$       &&
$250^{+44}_{-74}$   && $2.2$ &&
$5.7^{+1.0}_{-1.0}$    \\
A2256 && NCC (RCC) && $2.7^{+1.7}$         &&
$1.48^{+0.35}_{-0.29}$ && $6.2^{+3.9}_{-3.1}$     &&
$264^{+102}_{-80}$  && $2.2^{+0.3}_{-0.3}$  &&
$4.4^{+0.9}_{-0.9}$    \\
A644  && NCC (RCC) && $3.9_{-0.2}$ && $1.06^{+0.11}_{-0.11}$ &&
$0.7^{+0.1}_{-0.1}$     && $66^{+8}_{-9}$      &&
$2.1^{+0.4}_{-0.4}$  &&
$4.9^{+1.8}_{-1.8}$    \\
\enddata
\tablecomments{The classification is taken from Molendi \&
Pizzolato (2001), see also Henning et al. (2009). For A1689 and
A1656, the value of the virial radius is taken from the
literature (see text). The values of $k_B T_R$ are computed
from Eq.~(9), and are to be compared with those from the strong
shock condition (see \S~2 and 3). For A644 the values of
$\bar{k}_c$ and $r_f$ refer to the X-ray peak region; this
explains the low $k_c$ value (see Fig.~10).}
\end{deluxetable*}

Thus the CC clusters with their low value of $a$ appear to be
generally \emph{older} structures, currently in their stage of
slow and smooth accretion, with the main action taking place in
the outskirts under the form of calm entropy deposition by
gravitational accretion shocks. Toward the center, the CC
hallmark is constituted by a temperature peak
\emph{overlapping} (cf. Fig.~3 in CLFF09) the peak in
$\sigma^2$ at $r_m\approx 10^{-1}\, R$ (cf. the profiles of
A2199 and A2597 in Figs.~1-3). The condition for the peak to
occur after the SM is a low value of the central entropy
$\bar{k}_c \la 3\times 10^{-2}$, which comes to $30-50$ keV
cm$^2$. This behavior is highlighted in terms of $T(r) \propto
k(r)\, n(r)^{2/3}$; as the ICP density $n(r)$ rises
monotonically inward, $T(r)$ will peak and then decline toward
the center as $k(r)$ decreases sharply toward a low central
value $k_c$.

Such an inward decline of $T(r)$ to a low but finite central
value $T_c \propto k_c^{0.35}$ links to a high density $n_c
\propto k_c^{-1}$ to constitute the \emph{cool core}, a feature
of the non-radiative SM equilibrium (also present in
simulations discussed by Borgani et al. 2008). As expanded upon
by CLFF09, the SM does not include enhanced cooling, even less
any related inflow; it rather focuses the conditions for
enhanced radiation and fast cooling to set in on the timescale
$t_c\approx 0.3\,(k_c/15~\mathrm{keV}~\mathrm{cm}^2)^{1.2}$
Gyr. This would lead to a cooling catastrophe (e.g., White \&
Rees 1978; Blanchard et al. 1992), that may be stabilized by
ICP condensing around and into a central massive galaxy to
trigger accretion on the nuclear black hole. These conditions
kindle up AGN activities that drive rising bubbles or even
outgoing blastwaves, feed back entropy, and distribute it
widely into the ICP (see Binney \& Tabor 1995; Ciotti \&
Ostriker 2001; Cavaliere et al. 2002; Churazov et al. 2005;
Lapi et al. 2005; Voit \& Donahue 2005; Tucker et al. 2007).

We have analyzed in detail the two CC clusters A2199 and A2597
with their inward decrease of the temperature. We have found
\emph{low} central entropy levels $k_c\la 15$ keV cm$^{2}$
typical of CCs, with little or no need for an extended entropy
floor. We have derived outer powerlaw slopes $a\la 1$ (see
Table~1), lower than the standard value $1.1$ corresponding to
the standard concentration $c\approx 4$.

This trend culminates with A1689, a cluster with a CC-like
inner profile but featuring interesting \emph{outer}
peculiarities. On an empirical stand, our SM analysis confirms
the results by Lemze et al. (2008) concerning the high halo
concentration $c\approx 10$ (concurring with the gravitational
lensing analysis by Broadhurst et al. 2008; Lapi \& Cavaliere
2009b) and involving the \emph{flat} slope $a\approx 0.8$ for
the outer entropy profile. But we go beyond, and show in terms
Eqs.~(3) and (4) why these values deviate from the standard
ones $c\approx 4$ and $a\approx 1.1$, as spelled out above. In
the same vein, CLFF09 find that a \emph{steeper} density slope
$g = 3\,(a+b_R)/5\approx 2.4$ is to apply in the outskirts. The
high concentration of A1689 implies this to be an old structure
with the bulk region dating back to a transition epoch as early
as $z_t\approx 1.5$. In fact, the feature common to CC clusters
like A2199, A2597, and A1689 is constituted by low values of
$a\la 1$ and high values $c>4$ (see Table~1), that follow from
their being generally \emph{old} structures with shallow outer
potential wells.

At the other extreme, the NCC clusters appear to be dynamically
\emph{young} structures from our determination of DM
concentration and slope in the outer entropy profile. For
example, in A1656 (Coma Cluster) our SM fit requires a
\emph{high} value $a \approx 1.3$, and relatedly (see Eqs.~3
and 4) a \emph{low} concentration $c\approx 4$ and young age
$z_t \la 0.5$. Toward the center, the NCC clusters are marked
by a rising or flat temperature profile and by a generally flat
brightness distribution. This occurs for central levels of
$k_c$ \emph{exceeding} some $50$ keV cm$^{2}$, and also points
toward a thermodynamically young age for the ICP. In fact,
frequent and intense merger/AGN activity is expected in these
clusters observed in the aftermath of their fast initial
collapse, with considerable residual occurrence of mergers and
AGN outbursts that lead to large central injections of energy
and entropy.

Such features are exhibited, in a sequence of increasing
complexity, by A1656, A2256 and A644. Here, the SM elicits not
only a high level, but also a \emph{pattern} for the entropy
deposited in the form of a floor extended out to $r_f$. This we
interpret in terms of the stallation radius attained by a
powerful, outbound blast either triggered by a major head-on
merger (cf. simulations by Schindler et al. 2002, Vazza et al.
2009) or driven by a violent AGN outburst (see Forman et al.
2005; Cavaliere \& Lapi 2006; McNamara \& Nulsen 2007; Puchwein
et al. 2008), before being degraded into adiabatic sound waves
of the kind caught in action by Fabian et al. (2006) in the
Perseus cluster. To reach $r_f\approx 250$ kpc it takes a
rather extreme merger delivering about $10^{64}$ erg and
triggering a Sedov blastwave that expands as $R_s\propto
E^{1/3}\,t^{2/3}$ with decreasing Mach number; alternatively,
it takes an AGN outburst of about $10^{62}$ erg continuously
driving a blastwave to expand at constant Mach number with
$R_s\propto E^{1/3}\,t$, see Lapi et al. (2005) and Cavaliere
et al. (2006) for details. As discussed in \S~4.4 and 4.5, in
the NCC clusters A1656 and A2256 analyzed here such values of
$r_f$ are accompanied by evidence of ongoing mergers, hallmarks
of a recent cluster formation.

This interpretation relates $r_f$ to the \emph{dating} of the
merger responsible for the energy/entropy input; the good
performance of the SM implies such a time to be intermediate
between the blast transit time $r_f/ \mathcal{M}\, v_s \approx
10^{-1}$ Gyr (see Cavaliere \& Lapi 2006), and the time
$0.3\,(k_c/15~\mathrm{keV}~\mathrm{cm}^2)^{1.2}\approx 1$ Gyr
needed by radiative cooling to erode an entropy floor of about
$50$ keV cm$^2$.

Such a timing also guarantees that an accurate description of
the ICP thermodynamic state for both CC and NCC clusters is
provided by the SM based on the hydrostatic equilibrium
expressed by Eq.~(1). To complete the issue, note that not only
the equilibrium of the ICP is somewhat faster to attain than
the DM's (see Ricker \& Sarazin 2001; Lapi et al. 2005), but
also that circularized data (integrated over annuli, see
Snowden et al. 2008) tend to effectively smooth out local,
limited deviations from spherical hydrostatics and to better
agree with equilibrium.

How does the SM face the challenge of \emph{complexity} posed
by substructures as observed in A2256 and A644? Interestingly,
we still obtain good fits if we extrapolate the SM out to, or
perhaps beyond its literal limits, toward conditions where
$T(r)$ varies on the scale $r_f$, or differs around two
locations (cf. Figs.~8-11); these conditions highlight the
capabilities of the SM as a mere fitting tool. By the same
token, the SM provides sharp snapshots of physical conditions
even when these are spatially complex (as for A644), and
strongly suggests that they may be traced back to two merger
outcomes: a `hot spot' imprinted in A2256 to partially erase a
previous cool state and to yield a nascent NCC; or a `cold
drop' imported into the hot medium of A644 that will offset
cooling. Focusing on the cold component, such cases may be
termed as RCCs for Remnant of Cool Cores.

\begin{deluxetable}{cllllll}
\tabletypesize{} \tablecaption{Fit parameters of Eqs.~(A1) and
(A2)} \tablewidth{0pt} \tablehead{\colhead{$\alpha$} & &
\colhead{$s$} & & \colhead{$u$} & & \colhead{$q$}}\startdata
1.25 && 0.750 && 0.389 && 21.465\\
1.26 && 0.756 && 0.399 && 14.374\\
1.27 && 0.762 && 0.411 && 10.614\\
1.28 && 0.768 && 0.423 && 8.316\\
1.29 && 0.774 && 0.436 && 6.751\\
\enddata
\end{deluxetable}

In closing, we revert to Table~1 that provides an overall view
of the SM parameters for the present cluster set, to stress the
following points. First, columns 2, 3 and 4 visualize the
\emph{sharp} correlations among cluster classes (CC/NCC), basic
properties of the DM halo (high/low concentration implying
old/young dynamical age) and thermodynamic state of the ICP
(low/high central entropy level, narrow/wide entropy floor, and
steep/flat entropy slope). Second, columns 4 and 5 visualize
how from the density and temperature distributions over the
full data range we obtain entropy profiles of comparable
quality to Zhang et al. (2008) and Cavagnolo et al. (2009). In
particular, compared to the latter authors we find similar
values of $k_c$ and somewhat lower values of $a$ (with the
upper bound on $a$ discussed in \S~2); we note that our values
are derived by directly fitting the primary X-ray observables
$S_X(w)$ and $T(w)$ with the SM, rather than deprojected
profiles. Concerning virial radii, our values in column 7
systematically agree with literature evaluations; the
agreements of the related virial masses are commented upon in
\S~4.3, 4.4, 4.5. Finally, column 8 visualizes the boundary
values $k_B T_R$ useful to introduce a discussion of the actual
shock strengths related to the development of the outer halo
and to preheating levels of the (Warm-Hot) IGM (see \S~3).

To sum up, our analysis of several NCC and CC galaxy clusters
has shown how effective is the tool constituted by our
Supermodel. On using a simple formalism and a fast algorithm,
this leads to accurately fit in terms of a \emph{few} physical
parameters the \emph{many} data points concerning
\emph{diverse} temperature and brightness shapes to constitute
the library indexed by Table 1. Whence we extract sharp
information concerning the evolutionary stage and the thermal
history of clusters; in the inner regions these include the
\emph{level} ($k_c$), \emph{pattern} ($r_f$) and \emph{timing}
of the entropy injections into the ICP, and in the outskirts
the \emph{slope} ($a$) of the entropy deposited by accretion,
simply related to the halo \emph{concentration} ($c$) and
\emph{age} ($z_t$).

Finally, the Supermodel offers predictions as for the density
and temperature profiles in the cluster outskirts out to the
boundary $r\approx R$ facing the IGM; for example, within a
surrounding supercluster enhanced IGM preheating may be
expected to weaken the accretion shocks and lower the tail of
the brightness and temperature profiles. Such outer regions
challenge most current instruments but are coming of age with
\textsl{SUZAKU} (see Bautz et al. 2009; George et al. 2009;
Reiprich 2009), and will constitute a main target for the next
generation X-ray telescopes planned to study low surface
brightness plasmas, such as \textsl{WFXT} (see Giacconi et al.
2009; also \url{http://wfxt.pha.jhu.edu/}). These will open up
the way to use galaxy clusters as \emph{probes} of the
surrounding Warm-Hot IGM, an interesting perspective that we
will pursue in a following paper.

\textsc{Note added in proof}: The evidence concerning remnants
of cool cores in NCC clusters is also discussed from an
observational point of view by M. Rossetti \& S. Molendi (2009,
A\&A, submitted).

\begin{acknowledgments}
Work supported by ASI and INAF. We acknowledge stimulating
discussions with R. Giacconi, and informative exchanges with D.
Buote and F. Gastaldello. We thank our referee for insightful
comments, helpful toward improving our presentation. A.L.
thanks SISSA and INAF/OATS for hospitality.
\end{acknowledgments}

\begin{appendix}

\section{Analytic fits to the $\alpha$-profiles}

Here we provide analytic fits to the radial runs of both the
density and the circular velocity in the DM halos following the
(isotropic) $\alpha$-profiles.

It is convenient to express these in terms of the popular and
flexible parametric expression for generic density profiles
\begin{equation}
\bar{\rho}(\bar{r}) = {1\over \bar{r}^s}\,\left[{1+w\,c^u\over
1+w\,(c\,\bar{r})^u}\right]^q~
\end{equation}
introduced by Zhao (1996) and constituting an extension of the
empirical NFW (Navarro et al. 1997) formula (recall that
$\bar{r}=r/R$ is the adimensional radial coordinate, while
$c\equiv R/r_{-2}$ is the concentration parameter).

We cure the latter's unphysical features: diverging total mass
and steep central cusp corresponding to angled potential, by
deriving the parameters in Eq.~(A1) directly from the Jeans
equation and its derivatives, as shown by Lapi \& Cavaliere
(2009a, see their Appendix A). In fact, by differentiating
Eq.~(A1) it is easy to see that the latter approach implies
$w=-(2-s)/(2-s-q u)$. In addition, we find
\begin{equation}
s=\gamma_a~~~,~~~u={6\over 5}\,{\kappa_{\rm crit}\over
\gamma_0-\gamma_a}-{2\over
3}\,(\gamma_b-\gamma_a)~~~,~~~q={(\gamma_0-\gamma_a)^2\over
3\,\kappa_{\rm crit}/5-2\,(\gamma_0-\gamma_a)^2/3}~;
\end{equation}
here $\gamma_a=3\,\alpha/5$, $\gamma_0=6-3\,\alpha$,
$\gamma_b=3\,(1+\alpha)/2$, and $\kappa_{\rm
crit}(\alpha)\simeq 8.23-4.44\,\alpha$. In Table~2 we report
the parameter determinations for several values of $\alpha$.

Using Eq.~(A1), the function $\bar{v}_c^2(\bar{r})$ to be
inserted in Eq.~(2) of the main text may be explicitly
expressed as
\begin{equation}
\bar{v_c}^2(\bar{r}) = \bar{r}^{2-s}~~
{_2F_1[(3-s)/u~,q~,1+(3-s)/u~;-w\,(c\bar{r})^u]\over
_2F_1[(3-s)/u~,q~,1+(3-s)/u~;-w\,c^u]}~
\end{equation}
in terms of the hypergeometric function $_2F_1$; for a plot,
see Fig.~1 in CLFF09.

The resulting fits to the density and circular velocity
profiles of the $\alpha$-profiles hold to better than $15\%$ in
the whole range $10^{-2}\lesssim c\bar{r}\lesssim 10$. In
particular, for $\alpha=35/27=1.\overline{296}$ we recover the
exact solution of the Jeans equation found by Dehnen \&
McLaughlin (2005).

\end{appendix}

\end{document}